\documentclass{aa}
\usepackage{graphicx,lscape}
\usepackage{natbib}
\usepackage{color}
\usepackage{ulem}

\bibpunct{(}{)}{;}{a}{}{,} 
\usepackage{rotating,color}
\usepackage{supertabular}
\usepackage{longtable}
\usepackage{amssymb,color,natbib}
\usepackage{amsmath,amsfonts,amssymb,graphicx}

\begin{document}

\def\ms{$M_{\odot}$}

\newcommand{\red}{\textcolor[rgb]{1.0,0.0,0.0}}
\newcommand{\blue}{\textcolor[rgb]{0.0,0.0,0.7}}

\title{Chemical Evolution of Dwarf Irregular and Blue Compact Galaxies}

\author{J. Yin\inst{1,2}, F. Matteucci\inst{2,3}, G. Vladilo\inst{3}}

\authorrunning{J. Yin et al.}

\titlerunning{Chemical Evolution of dIrrs and BCDs}

\institute{ Key Laboratory for Research in Galaxies and Cosmology,
Shanghai Astronomical Observatory, CAS, 80 Nandan Road, Shanghai,
200030, China. \email{jyin@shao.ac.cn} \and  Department of Physics ,
Astronomy Division, Trieste University, Via G.B. Tiepolo 11, 34131
Trieste, Italy. \email{yin--matteucc@oats.inaf.it} \and Osservatorio
Astronomico di Trieste,INAF, Via G.B. Tiepolo 11, 34131 Trieste,
Italy. \email{vladilo@oats.inaf.it}}

\abstract{}{Dwarf irregular and blue compact galaxies are very
interesting objects since they are relatively simple and unevolved.
We aim at deriving the formation and chemical evolution history of
late-type dwarf galaxies, and compare it with DLA systems.} {We
present new models for the chemical evolution of these galaxies by
assuming different regimes of star formation (bursting and
continuous) and different kinds of galactic winds (normal and
metal-enhanced). The dark-to-baryonic mass ratio is assumed to be 10
in these models. The chemical evolution model follows in detail the
evolution of He, C, N, O, S, Si and Fe. We have collected the most
recent data on these galaxies and compared with our model results.
We have also collected data for Damped-Lyman $\alpha$-systems.} {Our
results show that in order to reproduce all the properties of these
galaxies, including the spread in the chemical abundances, the star
formation should have proceeded in bursts and the number of bursts
should be not larger than 10 in each galaxy, and that metal-enhanced
galactic winds are required. A metal-enhanced wind efficiency
increasing with galactic mass can by itself reproduce the observed
mass-metallicity relation although also an increasing efficiency of
star formation and/or number and/or duration of bursts can equally
well reproduce such a relation.} {Metal enhanced winds together with
an increasing amount of star formation with galactic mass are
required to explain most of the properties of these galaxies. Normal
galactic winds, where all the gas is lost at the same rate, do not
reproduce the features of these galaxies. On the other hand, a
global increase of the amount of star formation (increasing
efficiency and/or number of bursts and/or burst duration) with
galactic mass is able by itself to reproduce the mass-metallicity
relation even without winds, but without metal-enhanced winds is not
able to explain many other constraints. We suggest that these
galaxies should have suffered a different number of bursts varying
from 2 to 10 and that the efficiency of metal-enhanced winds should
have been not too high ($\lambda_{mw}\sim1$). We predict for these
galaxies present time Type Ia SN rates from 0.00084 and 0.0023 per
century. Finally, by comparing the abundance patterns of Damped
Lyman-$\alpha$ objects with our models we conclude that they are
very likely the progenitors of the present day dwarf irregulars.}

\keywords{Galaxies: evolution - Galaxies: abundance - Galaxies:
dwarf - Galaxies: irregular}

\maketitle

\section{Introduction}
Galaxy formation and evolution is one of the fundamental problems in
astrophysics. According to hierarchical clustering models, larger
galactic structures build up and grow through the accretion of dwarf
galaxies which are the first structures to collapse and form stars
\citep{White91, Kauffmann93}. These building-block galaxies are too
faint and small to be studied at high redshifts, while a class of
nearby metal-deficient dwarf galaxies offer a much better chance of
understanding it \citep{Thuan08}.

Dwarf galaxies, defined arbitrarily as galaxies having an absolute
magnitude fainter than $M_B\sim-18$ mag, are the most numerous
(about $80\%-90\%$) galaxies in the nearby universe \citep{Mateo98,
Grebel01, Karachentsev04}. Their space density has been suggested to
be about 40 times higher than that of bright galaxies
\citep{Staveley92}. Late-type dwarfs, dwarf irregular galaxies
(dIrrs) and blue compact dwarf galaxies (BCDs), are galaxies
harboring active or recent star formation activity, but have low
metallicities, large gas content and mostly young stellar
populations. All these features indicate that they are poorly
evolved objects, either newly formed galaxies or evolving slowly
over the Hubble time. Especially BCDs, the least chemically evolved
star-forming galaxies known in the universe (12+log(O/H) ranging
between 7.1 and $\sim$8.4), are excellent laboratories for studying
nucleosynthesis processes in a metal-deficient environment, in
conditions similar to those prevailing at the time of galaxy
formation \citep{Thuan95, Izotov99}. The study of the variations of
one chemical element relative to another in these poorly evolved
star-forming galaxies is crucial for our understanding of the early
chemical evolution of galaxies and for constraining models of
stellar nucleosynthesis.

DIrrs are dominated by scattered bright H{\sc ii} regions in the
optical, while in H{\sc i} they show a complicated fractal-like
pattern of shells, filaments and clumps. Typical H{\sc i} masses are
$\leq10^9M_{\odot}$.

BCDs are currently undergoing an intense burst of star formation
which gives birth to a large number ($10^3-10^4$) of massive stars
in a compact region( $\leq1$ kpc), which ionizes the interstellar
medium, producing high-excitation supergiant H{\sc ii} regions and
enriching it with heavy elements \citep{Thuan95}. Part of the
extended neutral gas may be kinematically decoupled from the
galaxies \citep{vanZee98}. The majority of BCDs (more than $99\%$)
are not primordial systems, but evolved dwarf galaxies where
starburst activity is immersed within an old extended stellar host
galaxy \citep{Kunth88, Papaderos96, Thuan08}. However, recent work
lends strong observational support to the idea that some among the
most metal-deficient star-forming galaxies known in the local
universe have formed most of their stellar mass within the last 1
Gyr, hence they qualify as young galaxy candidates
\citep{Papaderos02, Izotov04b, Pustilnik04, Aloisi07}.

\cite{Searle73} concluded that extremely blue galaxies should have
undergone intense bursts of star formation (SF) separated by long
quiescent periods (bursting SF). Recent detections of old underlying
stellar populations in most BCDs seem to corroborate their
suggestion and reveal at least another burst of SF besides the
present one, even in the case of the most metal-poor BCDs known, I
Zw 18 \citep{Ostlin00} and SBS 0335-052W \citep{Lipovetsky99}.

Aside from the bursting SF mode, gasping \citep{Tosi91} or mild
continuous \citep{Carigi99, Legrand00a, Legrand00b} SF regimes have
been proposed for dIrrs and BCDs. The gasping scenario, in which the
interburst periods are significantly shorter than the active phases,
is probably the more realistic picture for many of them
\citep{Schulte01}.

It is very likely that dIrrs and BCDs have suffered galactic winds.
In the past years, theorists have argued that winds carry heavy
elements out of galaxies, and that they remove a larger fraction of
the metals in lower mass galaxies \citep{Larson74, Dekel86,
DeYoung90, Mac99}.
The observational evidence of outflows from dwarf galaxies has grown
rapidly in time (e.g., \citealt{Meurer92, Martin96, Bomans97}). Only
recently, however, has it become possible to directly measure the
metal content of galactic winds and confirm that winds are indeed
metal enhanced (e.g., \citealt{Martin02}).

In the past years many models for the chemical evolution of these
galaxies appeared and tried to explain the intrinsic spread observed
in their properties \citep{Matteucci83, Matteucci85, Pilyugin93,
Marconi94, Bradamante98, Henry00, Lanfranchi03, Romano06}. Most of
these papers suggested that the spread can be reproduced by varying
the efficiency of star formation or galactic wind from galaxy to
galaxy or by assuming that there is self-pollution in the H{\sc ii}
regions where the abundances are measured \citep{Pilyugin93}.
Metal-enhanced winds with different prescriptions were studied
\citep{Marconi94, Bradamante98, Recchi01, Recchi04, Romano06}.

Besides the chemical abundances, the photometric and spectral
properties are also taken into account in some theoretical works
(e.g., \citealt{Vazquez03, Stasinska03, Martin08, Martin09}).
\citet{Martin08, Martin09} combined different codes of chemical
evolution, evolutionary population synthesis and photoionization,
and concluded that the closed box models with an attenuated bursting
SF and a initial star formation efficiency (SFE)
$\epsilon=0.1\sim0.3 $ can reproduce the observed abundances,
diagnostic diagrams and equivalent width-colour relations of local H
{\sc ii} galaxies.

In this work, we present a new series of chemical evolution models
for dIrrs and BCDs. The models are based on the original one of
\citet{Bradamante98}, but consider a larger number of chemical
species and updated stellar yields. We have tested both the bursting
and the continuous regime of star formation By comparing our model
results with the most recent data of dIrrs and Damped Lyman-$\alpha$
systems (DLAs), we aim at understanding the importance of galactic
winds, the history of star formation and the origin of the
mass-metallicity relation in dwarf irregulars. A comparison between
the properties of local dIrrs and those of high redshift DLAs will
allow us to understand the nature of DLAs and establish whether they
can be considered as the progenitors of local dIrrs and BCDs.
Moreover, by studying in detail the abundance patterns such as
[X/Fe] versus [Fe/H] will allow us to understand if these dwarf
galaxies can be the building blocks of more massive galaxies, as
suggested by the hierarchical clustering scenario of galaxy
formation.

This paper is organized as follows. In Section 2, we present the
observational constraints. In Section 3, the adopted chemical
evolution models are described. Our model results are presented in
details in Section 4 and discussed in Section 5.

\section{Observational Properties }

The metallicity, defined as the fraction of elements other than
hydrogen and helium by mass, is an important indicator of the
formation and evolutionary stage of a galaxy, and usually correlates
with macroscopic properties of late-type galaxies, e.g. luminosity,
mass, gas fraction, rotation speed, morphological type, etc. (e.g.,
\citealt{Garnett02, Pilyugin04, Lee06, Vaduvescu07}). Except
hydrogen and helium, oxygen is the most abundant element in the
universe and easy to be measured in H{\sc ii} regions because of its
bright emission lines. In practice, the oxygen abundance is usually
used to represent the metallicity of the galaxy.

\subsection{Luminosity-Metallicity ($L-Z$) relation and Mass-Metallicity ($M-Z$) relation}

\begin{figure}[!t] 
  \centering
  \includegraphics[width=0.45\textwidth]{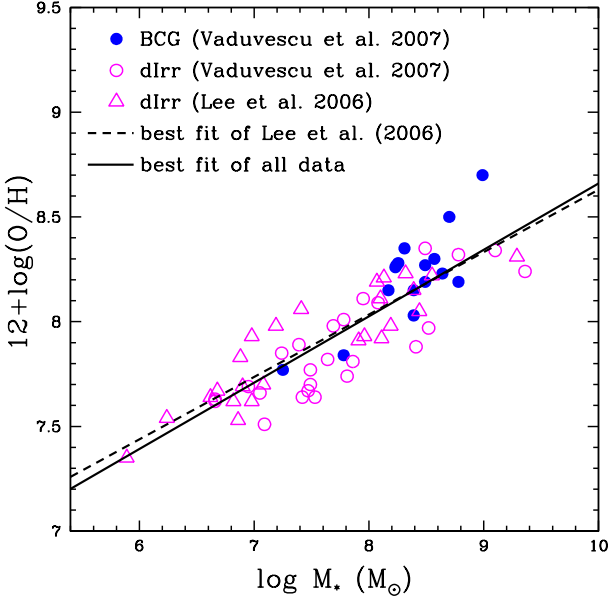}\\ 
  \caption{Oxygen abundance vs. stellar mass for nearby dIrrs and BCDs.
{\it Blue filled} and {\it magenta open circles} are data from
\cite{Vaduvescu07}, and represent BCDs and dIrrs respectively; {\it
magenta open triangles} are dIrrs observed by \cite{Lee06}, and the
{\it dashed line} shows the best fit of their data. The {\it solid
line} is the best linear fit of all data.}
  \label{Fig:obsMsZ}
\end{figure}

The strong correlation between the metallicity $Z$ and the
luminosity $L$ of a galaxy is a robust relationship, holding over 10
mag in galaxy optical luminosity and a factor of 100 in metallicity
(e.g., \citealt{Garnett87, Zaritsky94, Lamareille04, Tremonti04}).
\cite{Lequeux79} have shown first the existence of the correlation
between the metallicity and the mass in both compact and irregular
galaxies, then confirmed by \cite{Skillman89} who found that more
luminous (or more massive) galaxies are more metal rich. The
correlation is also found in spirals and elliptical galaxies (e.g.,
\citealt{Garnett87, Brodie91, Zaritsky94, vanZee97, Tremonti04,
Lee06, Vaduvescu07}). Since the luminosity of a galaxy closely
relates to its stellar mass, the $L-Z$ relation should represent
also the mass-metallicity relation. But how the luminosity is
representative of stellar mass it depends on the frequency band one
investigates. Traditionally, the $L-Z$ relation is studied at
optical wavelengths (e.g., \citealt{Lequeux79, Skillman89,
Skillman97, Pilyugin01, Garnett02, Lee03a, Lee03b, Pilyugin04,
vanZee06, Ekta10}). The optical luminosity could be affected by the
current star formation process, therefore more and more efforts were
put into the determination of the near-infrared $L-Z$ (and hence
$M_*-Z$) relation where the dominant emission arises from the older
stellar populations (e.g., \citealt{Perez03, Lee04, Salzer05, Lee06,
Mendes06, Rosenberg06, Vaduvescu07, Saviane08}). \cite{Lee06}
considered 27 nearby star-forming dwarf irregular galaxies whose
masses spread over 3 dex, and examined the $M_*-Z$ relation at 4.5
$\mu m$ (Spitzer)(see Fig. \ref{Fig:obsMsZ}). \cite{Vaduvescu05,
Vaduvescu06, Vaduvescu07} studied the properties of both dIrrs and
BCDs, and obtained the $M_*-Z$ relation by assuming
$M_*/L_K=0.8~M_{\odot}/L_{K\odot}$. They concluded that, for both
dIrrs and BCDs, metallicity correlates with stellar mass, gas mass,
and baryonic mass, in the sense that more massive systems are more
metal-rich.

\subsection{Gas fraction-metallicity ($\mu-Z$) relation }

\begin{figure}[!t]
  \centering
   \includegraphics[width=0.45\textwidth]{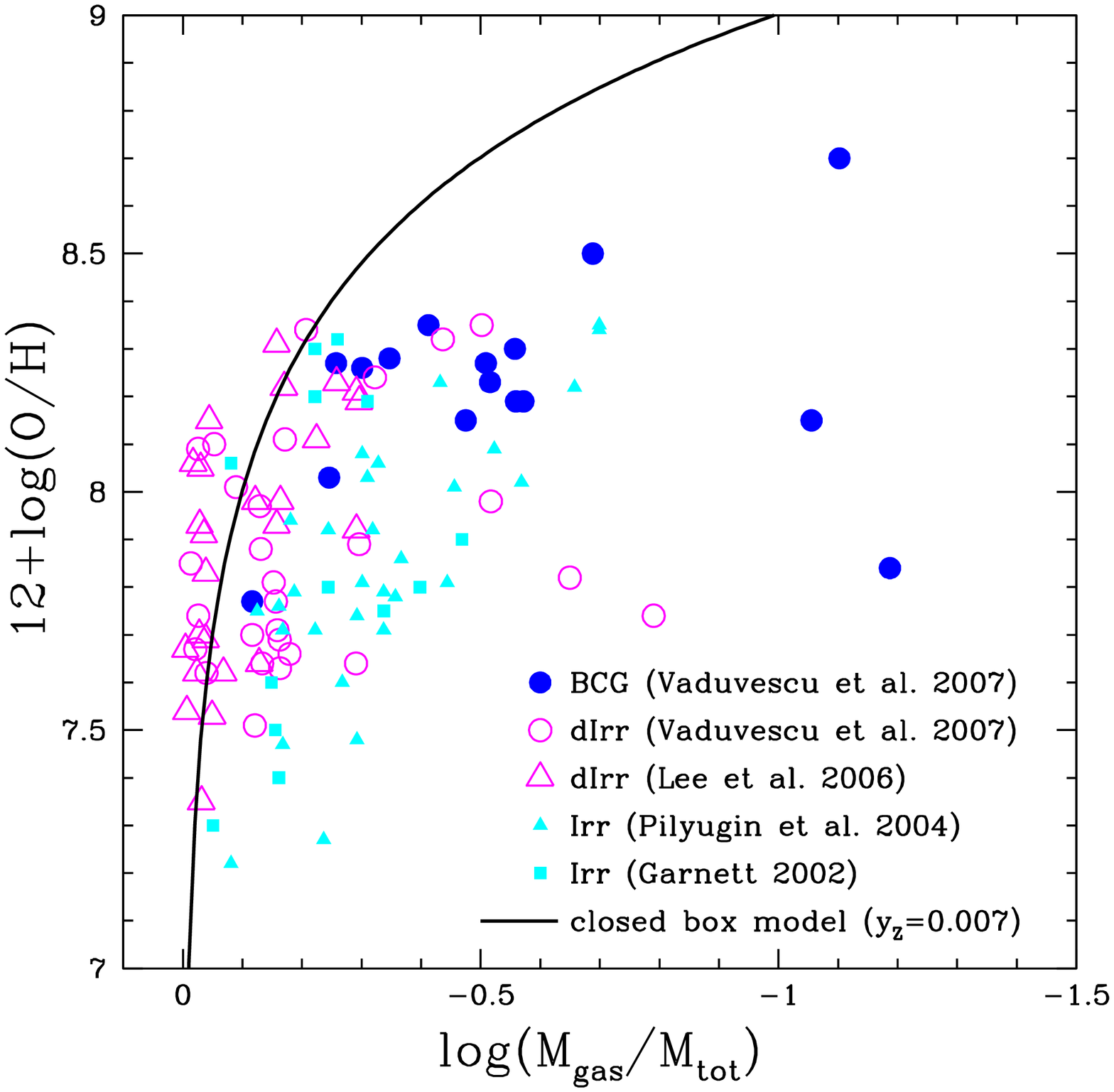}\\ 
  \caption{Oxygen abundance vs. gas fraction relation for nearby
dIrrs and BCDs. {\it Blue filled} and {\it magenta open circles} are
data from \cite{Vaduvescu07}, and represent BCDs and dIrrs
respectively; {\it magenta open triangles} are dIrrs observed by
\cite{Lee06}; {\it cyan filled triangles} and {\it squares} are the
Irrs observed by \cite{Pilyugin04} and \cite{Garnett02}
respectively. The {\it solid line} is the prediction of the closed
box model ($Z=y_Z\ln\mu^{-1}$) with an effective yield $y_Z=0.007$
corresponding to the Salpeter IMF and \citet{Woosley95}
nucleosynthesis.}
  \label{Fig:obsmuZ}
\end{figure}

A more useful relation for chemical evolution studies is the
metallicity-gas fraction ($\mu-Z$) relation, because it provides the
information about how the gas convert into stars and metals, and
also about gas flows (infall and/or outflow), as first suggested by
\cite{Matteucci83} in their chemical evolution models of dwarf
irregular galaxies.

As former works have shown, if the system does not have gas flows,
the metallicity evolution predicted by the closed box model is a
simple function of the gas fraction $\mu$ and true yield $y$ under
the instantaneous recycling assumptions, $Z=y_Z\ln(\mu^{-1})$
\citep{Schmidt63, Searle72}. By comparing the ``observed'' effective
yield, $y_{Z, eff}=Z_{obs}/\ln(\mu_{obs}^{-1})$ with the true
measured yield $y_Z$, one can understand whether the system evolved
as a closed box or if infall and/or outflow have been important. In
fact, both infall and outflow have the effect of decreasing the
effective yield.

\begin{figure*}[!t]
  \centering
   \includegraphics[width=0.8\textwidth]{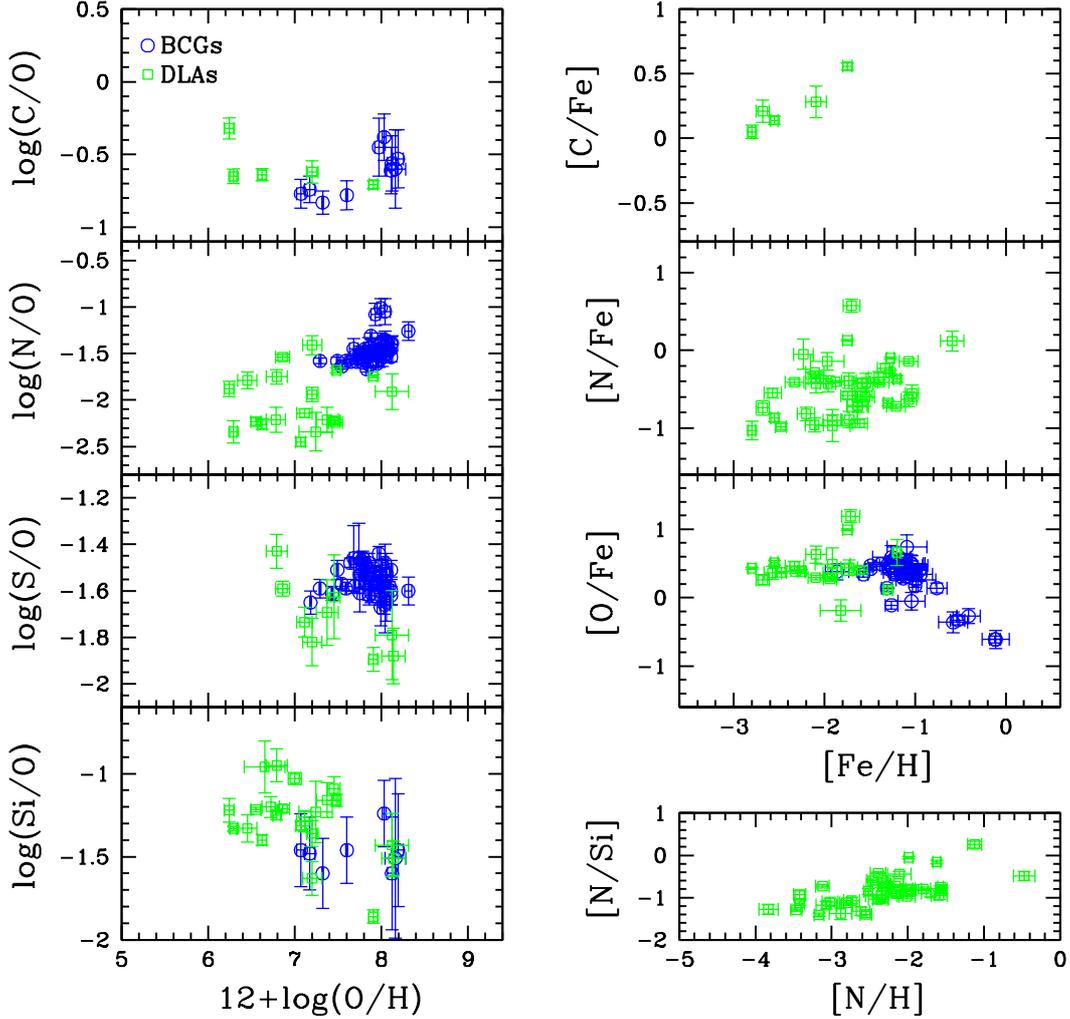}\\
  \caption{Abundance ratio of different elements. {\it Blue open
  circles} are BCDs, from \cite{Izotov99} and \cite{Papaderos06};
  {\it green open squares} are DLAs, the data are shown in Table~\ref{Tab:DLA}.}
  \label{Fig:obsabund}
\end{figure*}

\cite{Lee03a, Lee03b} showed that the oxygen abundance is tightly
correlated with the gas fraction in dwarf irregular galaxies through
optical observations, and they argued that these dIrrs have evolved
in relative isolation, without inflow or outflow of gas. Lately,
\cite{Lee06} measured the 4.5 $\mu m$ luminosities for 27 nearby
dIrrs with the $Spitzer$ Infrared Array Camera, and showed the
relation between oxygen abundances and gas-to-stellar mass ratio.
Their results suggest reduced yields and/or significant outflow
rates, which have been also indicated by previous authors (e.g.,
\citealt{Garnett02, vanZee06}). Using NIR photometry, the most
important discovery of \cite{Vaduvescu07} is that the $\mu-Z$
relation for BCDs follows that of dIrrs, and agrees with
\cite{Lee03a, Lee03b} that the evolution of field dIrrs has not been
noticeably influenced by gas flows. However, they also point out
that several dIrrs and at least one BCD do show HI deficiencies in
dense environments, indicating that the gas may be removed by
external processes. Using the effective yield as the only criterion
for gas flow is too simplistic and the conclusion might not be
reliable. Convincing  conclusions should be reached through detailed
modeling work.

In Fig.~\ref{Fig:obsmuZ}, we show the observed oxygen abundance-gas
fraction relation of nearby dIrrs and BCDs.
>From this figure we can see, the BCDs have larger abundance and gas
fraction range, while most dIrrs have higher gas fraction compared
with BCDs and Irrs, implying their poorly evolved stages.

\subsection{Abundance ratios }


\begin{sidewaystable*}\tiny
\vspace{18cm}
  \centering
  \caption{\normalsize The abundances of DLAs }\label{Tab:DLA}
\begin{tabular}{lcllclccccccl}
\hline \hline
 QSO name    &$z_{abs}$&dv(km/s)&  ref. &  log$N_{HI}$       & ref. &  [C/H]              &  [N/H]                &  [O/H]                &  [Si/H]               &  [S/H]                &  [Fe/H]               &    ref.  \\
\hline
 1331+170    & 1.7764  &   75  &   1    &  $ 21.14\pm0.08 $  & 2    &                     &    $ -2.390\pm0.128 $ &                       &    $ -1.350\pm0.081 $ &                       &    $ -1.960\pm0.085 $ &    2,3    \\
 2230+025    & 1.8642  &   59  &   4    &  $ 20.83\pm0.05 $  & 5    &                     &    $ -1.588\pm0.105 $ &                       &    $ -0.641\pm0.071 $ &                       &    $ -1.035\pm0.071 $ &    5    \\
 2314-409    & 1.8750  &   59  &   4    &  $ 20.10\pm0.20 $  & 5    &                     &                       &    $ -2.010\pm0.233 $ &    $ -1.820\pm0.224 $ &                       &    $ -1.820\pm0.224 $ &    5    \\
 1210+1731   & 1.8918  &   62  &   1    &  $ 20.63\pm0.08 $  & 6    &                     &    $ -1.700\pm0.120 $ &                       &    $ -0.810\pm0.085 $ &                       &    $ -1.070\pm0.085 $ &    7    \\
 2206-199A   & 2.0762  &   20  &   1    &  $ 20.43\pm0.04 $  & 8    &  $ -2.410\pm0.050 $ &    $ -3.420\pm0.064 $ &    $ -2.040\pm0.050 $ &    $ -2.290\pm0.041 $ &                       &    $ -2.550\pm0.041 $ &    8    \\
 1444+014    & 2.0870  &   294 &   1    &  $ 20.25\pm0.07 $  & 1    &                     &    $ -1.124\pm0.092 $ &                       &    $ -1.380\pm0.092 $ &                       &    $ -1.700\pm0.092 $ &    9            \\
 1037-2703   & 2.1390  &       &        &  $ 19.70\pm0.10 $  & 10   &                     &    $ -0.470\pm0.141 $ &                       &    $ 0.020 \pm0.102 $ &                       &    $ -0.590\pm0.128 $ &    10    \\
 0528-2505   & 2.1410  &   105 &   1    &  $ 20.95\pm0.05 $  & 11   &                     &    $ -2.150\pm0.094 $ &                       &    $ -1.240\pm0.071 $ &                       &    $ -1.550\pm0.103 $ &    11            \\
 2348-147    & 2.2790  &   55  &   1    &  $ 20.59\pm0.08 $  & 7    &                     &    $ -3.021\pm0.106 $ &                       &    $ -1.921\pm0.094 $ &                       &    $ -2.203\pm0.094 $ &    6    \\
 2036-0553   & 2.2803  &       &        &  $ 21.20\pm0.15 $  & 12   &                     &    $ -2.110\pm0.153 $ &                       &    $ -1.670\pm0.158 $ &                       &    $ -1.970\pm0.186 $ &    12   \\
 0100+130    & 2.3090  &   37  &   1    &  $ 21.37\pm0.08 $  & 2    &                     &    $ -2.120\pm0.128 $ &                       &                       &                       &    $ -1.730\pm0.081 $ &    2    \\
 2243-6031   & 2.3300  &   173 &   1    &  $ 20.67\pm0.02 $  & 13   &                     &    $ -1.570\pm0.036 $ &    $ -0.540\pm0.191 $ &    $ -0.820\pm0.028 $ &    $ -0.810\pm0.036 $ &    $ -1.200\pm0.036 $ &    13       \\
 1232+0815   & 2.3377  &   85  &   1    &  $ 20.80\pm0.10 $  & 11   &                     &    $ -1.950\pm0.128 $ &                       &    $ -1.130\pm0.135 $ &                       &    $ -1.540\pm0.128 $ &    11   \\
 1435+5359   & 2.3427  &       &        &  $ 21.05\pm0.10 $  & 12   &                     &    $ -2.160\pm0.102 $ &                       &    $ -1.430\pm0.102 $ &                       &                       &    12  \\
 0841+129    & 2.3745  &   37  &   1    &  $ 21.00\pm0.10 $  & 11   &                     &    $ -2.160\pm0.104 $ &                       &    $ -1.302\pm0.108 $ &                       &    $ -1.580\pm0.108 $ &    7,11   \\
 0027-1836   & 2.4020  &       &        &  $ 21.75\pm0.10 $  & 14   &                     &    $ -2.280\pm0.215 $ &                       &    $ -1.590\pm0.128 $ &                       &    $ -2.230\pm0.108 $ &    14   \\
 0112-306    & 2.4191  &   31  &   1    &  $ 20.50\pm0.08 $  & 15   &                     &    $ -3.120\pm0.089 $ &    $ -2.210\pm0.113 $ &    $ -2.389\pm0.082 $ &                       &    $ -2.570\pm0.094 $ &    9,15,16   \\
 2343+1232   & 2.4313  &   289 &   1    &  $ 20.40\pm0.07 $  & 17   &                     &    $ -1.560\pm0.076 $ &                       &    $ -0.760\pm0.092 $ &                       &    $ -1.330\pm0.086 $ &    17  \\
 1409+0930   & 2.4562  &   69  &   1    &  $ 20.54\pm0.04 $  & 18   &                     &                       &    $ -1.870\pm0.045 $ &    $ -1.970\pm0.045 $ &                       &    $ -2.250\pm0.045 $ &    16,18   \\
 1223+178    & 2.4661  &   91  &   1    &  $ 21.40\pm0.10 $  & 15   &                     &    $ -2.350\pm0.205 $ &                       &    $ -1.406\pm0.104 $ &                       &    $ -1.640\pm0.112 $ &    9,15,19   \\
 0841+129    & 2.4764  &   30  &   1    &  $ 20.78\pm0.08 $  & 6    &                     &    $ -2.618\pm0.120 $ &    $ -1.287\pm0.128 $ &    $ -1.296\pm0.085 $ &    $ -1.460\pm0.128 $ &    $ -1.726\pm0.094 $ &    6    \\
 1337+1121   & 2.5079  &   32  &   1    &  $ 20.12\pm0.05 $  & 16   &                     &    $ -3.060\pm0.112 $ &    $ -1.880\pm0.112 $ &                       &                       &                       &    16    \\
 2344+12     & 2.5379  &   69  &   1    &  $ 20.32\pm0.07 $  & 20   &                     &    $ -2.321\pm0.076 $ &                       &    $ -1.580\pm0.076 $ &                       &    $ -1.740\pm0.077 $ &    1,19   \\
 0405-443    & 2.5505  &   165 &   1    &  $ 21.13\pm0.10 $  & 21   &                     &    $ -2.360\pm0.104 $ &                       &    $ -1.320\pm0.108 $ &                       &    $ -1.630\pm0.117 $ &    21            \\
 1558+4053   & 2.5533  &       &        &  $ 20.30\pm0.04 $  & 8    &  $ -2.470\pm0.072 $ &    $ -3.420\pm0.081 $ &    $ -2.420\pm0.057 $ &    $ -2.490\pm0.072 $ &                       &    $ -2.680\pm0.072 $ &    8    \\
 0405-443    & 2.5950  &   79  &   1    &  $ 21.09\pm0.10 $  & 21   &                     &    $ -1.800\pm0.102 $ &                       &    $ -1.010\pm0.104 $ &                       &    $ -1.390\pm0.102 $ &    21   \\
 0913+072    & 2.6184  &   22  &   1    &  $ 20.34\pm0.04 $  & 8    &  $ -2.750\pm0.064 $ &    $ -3.830\pm0.126 $ &    $ -2.370\pm0.041 $ &    $ -2.550\pm0.041 $ &                       &    $ -2.800\pm0.041 $ &    8    \\
 0405-443    & 2.6215  &   182 &   1    &  $ 20.47\pm0.10 $  & 21   &                     &                       &    $ -1.940\pm0.102 $ &    $ -1.990\pm0.117 $ &                       &    $ -2.320\pm0.102 $ &    21             \\
 1759+7539   & 2.6250  &   74  &   22   &  $ 20.76\pm0.01 $  & 23   &                     &    $ -1.555\pm0.027 $ &                       &    $ -0.720\pm0.061 $ &                       &    $ -1.274\pm0.014 $ &    23,24   \\
 0812+32     & 2.6260  &   70  &   22   &  $ 21.35\pm0.10 $  & 25   &                     &                       &    $ -0.520\pm0.135 $ &    $ -0.880\pm0.112 $ &    $ -0.880\pm0.128 $ &    $ -1.710\pm0.100 $ &    26   \\
 1409+0930   & 2.6682  &   36  &   4    &  $ 19.70\pm0.04 $  & 18   &                     &    $ -1.980\pm0.045 $ &    $ -1.180\pm0.045 $ &    $ -1.190\pm0.050 $ &                       &    $ -1.300\pm0.050 $ &    16,18   \\
 1558-0031   & 2.7026  &       &        &  $ 20.67\pm0.05 $  & 27   &                     &    $ -1.990\pm0.054 $ &    $ -1.460\pm0.112 $ &    $ -1.940\pm0.054 $ &    $ -1.760\pm0.054 $ &                       &    8,12    \\
 1337+1121   & 2.7957  &   42  &   1    &  $ 20.95\pm0.10 $  & 25   &                     &    $ -2.740\pm0.104 $ &    $ -1.870\pm0.122 $ &    $ -1.670\pm0.122 $ &    $ -1.780\pm0.102 $ &    $ -2.330\pm0.102 $ &    16,26   \\
 1426+6039   & 2.8268  &   136 &   22   &  $ 20.30\pm0.04 $  & 28   &                     &    $ -1.370\pm0.041 $ &                       &                       &                       &    $ -1.270\pm0.041 $ &    26             \\
 1946+7658   & 2.8443  &   22  &   4    &  $ 20.27\pm0.06 $  & 28   &                     &    $ -3.462\pm0.072 $ &    $ -2.110\pm0.061 $ &    $ -2.176\pm0.061 $ &                       &    $ -2.478\pm0.061 $ &    19   \\
 2342+3417   & 2.9082  &   100 &   22   &  $ 21.15\pm0.10 $  & 25   &                     &    $ -2.010\pm0.108 $ &                       &    $ -1.040\pm0.102 $ &                       &    $ -1.580\pm0.117 $ &    26   \\
 1021+3001   & 2.9490  &   70  &   22   &  $ 20.70\pm0.10 $  & 25   &                     &    $ -3.070\pm0.135 $ &                       &    $ -1.890\pm0.102 $ &                       &    $ -2.110\pm0.100 $ &    26   \\
 0001        & 3.0000  &   75  &   4    &  $ 20.70\pm0.05 $  & 19   &                     &    $ -3.165\pm0.064 $ &    $ -1.594\pm0.055 $ &    $ -1.758\pm0.051 $ &                       &                       &    19   \\
 0741+4741   & 3.0174  &   42  &   22   &  $ 20.48\pm0.10 $  & 26   &                     &    $ -2.283\pm0.100 $ &                       &    $ -1.636\pm0.100 $ &                       &    $ -1.878\pm0.100 $ &    19   \\
 0347-383    & 3.0250  &   93  &   1    &  $ 20.73\pm0.05 $  & 1    &  $ -1.191\pm0.055 $ &    $ -1.620\pm0.051 $ &    $ -0.754\pm0.051 $ &    $ -1.464\pm0.064 $ &    $ -1.128\pm0.071 $ &    $ -1.748\pm0.051 $ &    9,15,29   \\
 2332-0924   & 3.0572  &   111 &   1    &  $ 20.50\pm0.07 $  & 16   &                     &    $ -2.550\pm0.076 $ &    $ -1.210\pm0.073 $ &    $ -1.150\pm0.099 $ &    $ -1.316\pm0.193 $ &    $ -1.610\pm0.086 $ &    9,16,15,25            \\
 2059-360    & 3.0830  &   44  &   1    &  $ 20.98\pm0.08 $  & 15   &                     &    $ -2.810\pm0.082 $ &    $ -1.550\pm0.089 $ &    $ -1.687\pm0.094 $ &    $ -1.764\pm0.094 $ &    $ -1.910\pm0.106 $ &    15,16,30   \\
 1340+136    & 3.1180  &   153 &   1    &  $ 20.05\pm0.08 $  & 16   &                     &    $ -2.550\pm0.082 $ &    $ -1.190\pm0.082 $ &                       &                       &                       &    16               \\
 0930+2858   & 3.2350  &   26  &   22   &  $ 20.35\pm0.10 $  & 25   &                     &    $ -2.390\pm0.101 $ &                       &    $ -1.972\pm0.102 $ &                       &    $ -2.105\pm0.101 $ &    19   \\
 0900+4215   & 3.2458  &   95  &   4    &  $ 20.30\pm0.10 $  & 26   &                     &    $ -1.930\pm0.102 $ &                       &                       &                       &    $ -1.210\pm0.102 $ &    12,26   \\
 0201+1120   & 3.3848  &   67  &   22   &  $ 21.30\pm0.10 $  & 31   &                     &    $ -1.750\pm0.117 $ &                       &                       &                       &    $ -1.400\pm0.113 $ &    19,32   \\
 0000-263    & 3.3901  &   33  &   1    &  $ 21.41\pm0.08 $  & 28   &                     &    $ -2.460\pm0.085 $ &    $ -1.800\pm0.081 $ &    $ -1.860\pm0.082 $ &    $ -1.870\pm0.085 $ &    $ -2.095\pm0.085 $ &    28,33   \\
 1108-0747   & 3.6080  &   31  &   1    &  $ 20.37\pm0.07 $  & 1    &                     &                       &    $ -1.660\pm0.076 $ &    $ -1.540\pm0.073 $ &                       &    $ -1.936\pm0.071 $ &    1,16,26   \\
 1443+2724   & 4.2240  &   130 &   1    &  $ 20.95\pm0.10 $  & 1    &                     &    $ -1.210\pm0.100 $ &                       &                       &                       &    $ -1.070\pm0.104 $ &    1              \\
 1202-0725   & 4.3829  &   170 &   22   &  $ 20.55\pm0.03 $  & 34   &  $ -1.810\pm0.058 $ &    $ -2.520\pm0.058 $ &    $ -1.460\pm0.067 $ &    $ -1.670\pm0.058 $ &                       &    $ -2.092\pm0.114 $ &    28,34            \\
 0307-4945   & 4.4660  &   192 &   22   &  $ 20.67\pm0.09 $  & 35   &                     &    $ -2.880\pm0.150 $ &    $ -1.420\pm0.192 $ &    $ -1.500\pm0.114 $ &                       &    $ -1.910\pm0.192 $ &    35   \\
\hline \hline
\end{tabular} \\
\tablebib{(1)\citet{Ledoux06}; (2)\citet{Dessauges04};
(3)\citet{Kulkarni96}; (4)\citet{Abate08}; (5)\citet{Ellison01a};
(6)\citet{Dessauges06}; (7)\citet{Dessauges07};
(8)\citet{Pettini08}; (9)\citet{Ledoux03}; (10)\citet{Srianand01};
(11)\citet{Centurion03}; (12)\citet{Henry07}; (13)\citet{Lopez02};
(14)\citet{Noterdaeme08}; (15)\citet{Srianand05};
(16)\citet{Petitjean08}; (17)\citet{Noterdaeme07};
(18)\citet{Pettini02}; (19)\citet{Prochaska02a};
(20)\citet{Peroux06}; (21)\citet{Lopez03}; (22)\citet{Prochaska08};
(23)\citet{Outram99}; (24)\citet{Prochaska02b};
(25)\citet{Prochaska03}; (26)\citet{Prochaska07};
(27)\citet{OMeara06}; (28)\citet{Lu96}; (29)\citet{Levshakov02};
(30)\citet{Petitjean00}; (31)\citet{Storrie00};
(32)\citet{Ellison01b}; (33)\citet{Molaro01};
(34)\citet{DOdorico04}; (35)\citet{Dessauges01}.}
\end{sidewaystable*}

How the abundances of chemical elements change relative to one
another is a crucial clue for understanding the chemical evolution
of galaxies and stellar nucleosynthesis.

H{\sc ii} regions are ionized by newly born massive stars, hence
showing the metallicity of the ISM at the present time. Therefore,
metallicities in dIrrs and BCDs are usually derived from the ionized
gas in H{\sc ii} regions through their strong narrow emission lines.
\cite{Izotov99} presented high-quality ground-based spectroscopic
observations of 54 supergiant H{\sc ii} regions in 50
low-metallicity BCDs with oxygen abundances 12+log(O/H) between 7.1
and 8.3, and determined abundances for the elements N, O, Ne, S, Ar,
Fe, and also C and Si in a subsample of 7 BCDs. \cite{Papaderos06}
presented spectroscopic and photometric studies of nearby BCDs in
the 2dFGRS (Two-Degree Field Galaxy Redshift Survey), and measured
their Ne/O, Fe/O and Ar/O ratios. Both of these works do not
consider the dust depletion correction. We show the data of
\cite{Izotov99} and \cite{Papaderos06} in Fig.~\ref{Fig:obsabund}.

\begin{figure*}[!tb]
  \centering
   \includegraphics[width=0.8\textwidth]{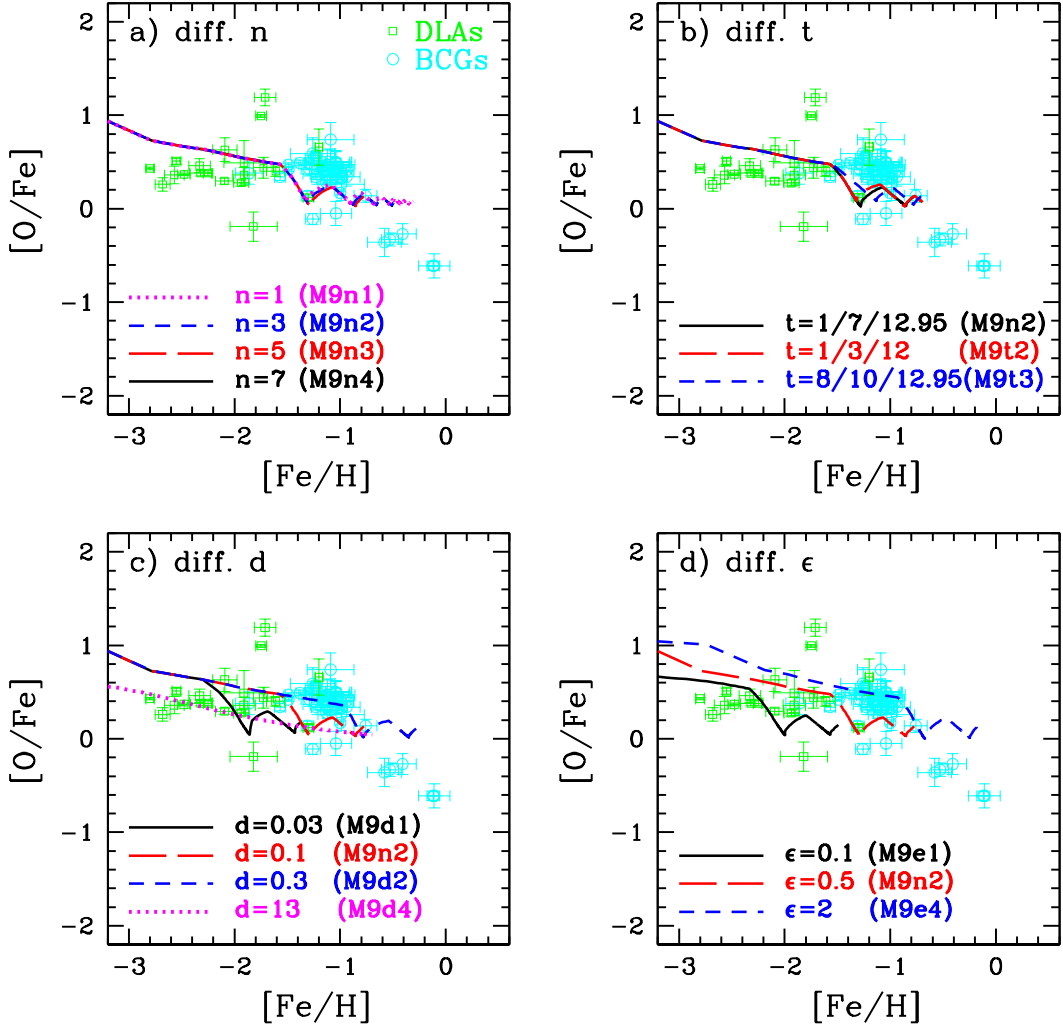}\\
  \caption{The evolutionary tracks of [O/Fe] vs. [Fe/H] as
  predicted by models without outflow.
  a) models with different number of bursts;
  b) models with different occurrence times of bursts;
  c) models with different durations of bursts;
  d) models with different star formation efficiency.
   All these models assume the same total infall mass $M_{inf}=10^9$\ms.
  The {\it cyan open circles} are BCDs
  and {\it green open squares} are DLAs,
  same as in Fig.~\ref{Fig:obsabund}.  }
  \label{Fig:nowdOFeH}
\end{figure*}

DLA absorption systems, found in the spectra of high-redshift QSOs,
are neutral clouds with large H{\sc i} column densities, $N({\rm
HI})\geq2\cdot10^{20}{\rm cm}^{-2}$. They are likely to be
protogalactic clumps embedded in dark matter halos and may provide
the important information on the early chemical evolution of
galaxies. With high resolution spectroscopy of QSO absorption lines,
elemental abundances can be measured up to redshift $z\approx5$.
The chemical abundances of DLA systems give us complementary
observational constraints on the formation and evolution of
galaxies. Abundance measurements in DLA systems relevant for the
present work are listed in Table~\ref{Tab:DLA} and plotted in
Fig.~\ref{Fig:obsabund} (green squares).
In comparing DLA abundances with model predictions care must be
taken for dust depletion effects. Luckily, these effects are
expected to be negligible for most of the elements used in the
present investigation, such as C, N, O and S: in fact these elements
show little values of depletion, if any, in nearby interstellar
clouds \citep{Jenkins09} and are expected to be even less depleted
in DLA systems. On the other hand, we expect some depletion effects
for Fe and, to a lesser extent, for Si. Estimates of Fe depletion in
DLAs based on the comparison with Zn measurements \citep{Vladilo04}
are available only for a few systems of Table~\ref{Tab:DLA}. These
results indicate that Fe tend to be underestimated when the level of
metallicity is relatively high. This explain the few cases with
largest deviations from BCD measurements and from the model
predictions shown in Figs. \ref{Fig:obsabund}, \ref{Fig:nowdOFeH},
\ref{Fig:wdZmuY}, \ref{Fig:mwdZmuY1}, \ref{Fig:mwdZmuY2},
\ref{Fig:bestabund}, and \ref{Fig:DLAabund}.

\subsection{Primordial helium abundance $Y_p$ and $\bigtriangleup Y$/$\bigtriangleup Z$ }

The determination of primordial helium abundance, $Y_p$, is
important for the study of cosmology and the evolution of galaxies,
because an accurate initial $Y$ is required to test Big Bang
nucleosynthesis and build chemical evolution models.

One way of estimating $Y_p$ is by extrapolating the observed
helium-metallicity ($Y-Z$) relation  to $Z=0$ by assuming the slope
$\bigtriangleup Y$/$\bigtriangleup Z$ to be  constant. More
recently, it has been common practice to use $\bigtriangleup
Y$/$\bigtriangleup O$ since the oxygen abundance is easier to
determine and can represent the metals. To obtain an accurate $Y_p$
value, a reliable determination of $\bigtriangleup
Y$/$\bigtriangleup O$ for oxygen-poor objects is needed (e.g.,
\citealt{Izotov99b, Peimbert03, Luridiana03, Izotov04a,
Peimbert07}).  \cite{Izotov04a} derived the primordial helium
$Y_p=0.2429\pm0.0009$ and the slope $\bigtriangleup
Y$/$\bigtriangleup O=4.3\pm0.7$ from observations of 82 H{\sc ii}
regions. For a restricted sample (7 H{\sc ii} regions), they
obtained $Y_p=0.2421\pm0.0021$ and $\bigtriangleup
Y$/$\bigtriangleup O=5.7\pm1.8$. Later, \cite{Izotov06} derived
$Y_p=0.2463\pm0.0030$ from the emission of the whole H{\sc ii}
region of the extremely metal-deficient blue compact dwarf galaxy
SBS $0335-052$E. \cite{Peimbert07} has adopted $\bigtriangleup
Y$/$\bigtriangleup O=3.3\pm0.7$ from theoretical and observational
results, and derived $Y_p=0.2474\pm0.0029$. These values are in
excellent agreement with the value derived by \cite{Spergel07} from
the WMAP results, $Y_p=0.2482\pm0.0004$.

\begin{figure}[!t]
  \centering
   \includegraphics[width=0.45\textwidth]{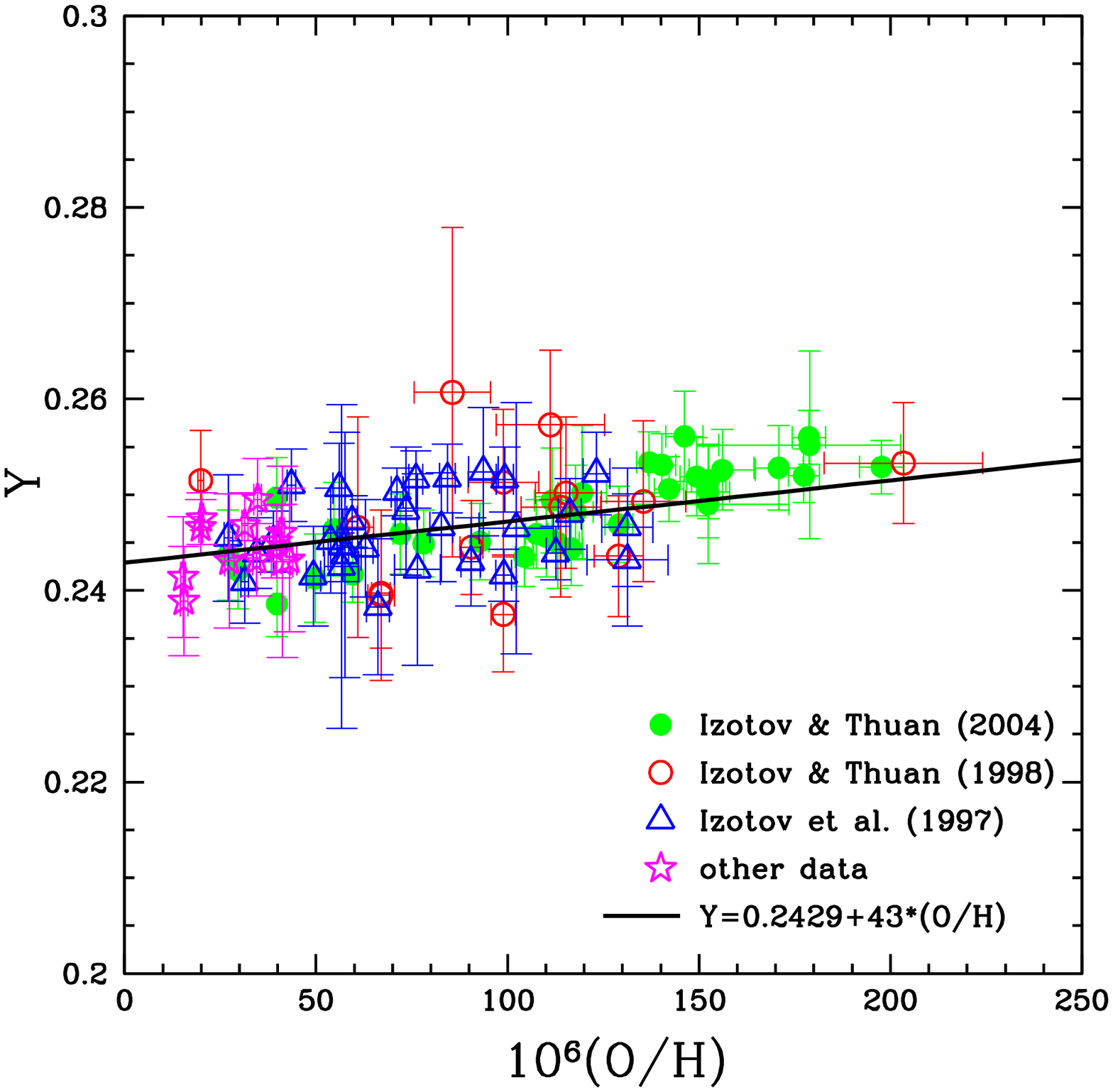}\\
  \caption{Helium-metallicity ($Y-Z$) relation for H{\sc ii} regions
in BCDs, Fig. 2 of \cite{Izotov04a}. {\it Green filled circles},
from \cite{Izotov04a}; {\it red open circles}, from
\cite{Izotov98b}; {\it blue open triangles}, from \cite{Izotov97};
{\it magenta open pentagrams}, from other data \citep{Izotov98a,
Izotov99b, Thuan99, Izotov01a, Izotov01b, Guseva01, Guseva03a,
Guseva03b}; the {\it solid line} is the  maximum likelihood linear
regression of all data, $Y=0.2429+43*(O/H)$}.
  \label{Fig:obsYZ}
\end{figure}

In Fig.~\ref{Fig:obsYZ} we replot the helium-oxygen abundance
relation of \cite{Izotov04a} by using their data in Table 5.
 The linear regression is the one derived from the whole sample
$Y=0.2429+43*(O/H)$.

\section{Model Prescriptions}

In this work, we used an updated version of the chemical evolution
model developed by \cite{Bradamante98} to study the formation and
evolution of late-type dwarf galaxies, dIrrs and BCDs.

The general picture is the following: our model is one-zone and
assumes the galaxy built up by continuous infall of primordial gas
($X=0.7571,~Y_p=0.2429,~Z=0$). Stars form  and then contaminate the
interstellar medium (ISM) with their newly produced elements which
mix with the ISM instantaneously and completely. Stellar lifetimes
are taken into account in detail, i.e. the instantaneous recycling
approximation (IRA) is relaxed. The energy released by supernovae
(SNe) and stellar winds is partially deposited in the ISM, and
galactic winds develop when the thermal energy of the gas exceeds
its binding energy. The wind expels metals from the galaxy, hence it
has a significant influence on the chemical enrichment of the
galaxy.

The time evolution of the fractional mass of the element $i$ in the
gas, $G_i$, is described by the equations:
\begin{equation}\label{eq:Gi}
\dot{G}_i(t)=-\psi(t)X_i(t)+R_i(t)+\dot{G}_{i,inf}(t)-\dot{G}_{i,out}(t),
\end{equation}
where $G_i(t)=M_g(t)X_i(t)/M_L(t_G)$ is the gas mass in the form of
an element $i$ normalized to the total baryonic mass $M_L$ at the
present day $t_G=13$~Gyr; $M_g(t)$ is the gas mass at time $t$ and
$X_i(t)$ represents the mass fraction of element $i$ in the gas,
i.e., abundance by mass. The quantity $G(t)=M_g(t)/M_L(t_G)$
represents the total fractional mass of gas and $X_i(t)$ can be
expressed by $G_i(t)/G(t)$. The four items on the right hand side of
equation (\ref{eq:Gi}) show the mass change of the element $i$
caused by the formation of new stars $\psi(t)X_i(t)$, the material
returned through stellar winds or SN explosion $R_i(t)$, the infall
of primordial gas $\dot{G}_{i,inf}(t)$, and the outflow
$\dot{G}_{i,out}(t)$ respectively.

The star formation rate (SFR) $\psi(t)$ in this work is simply
assumed as:
\begin{equation}\label{eq:psi}
    \psi(t)=\epsilon G(t),
\end{equation}
where $\epsilon$ is the star formation efficiency and is in units of
Gyr$^{-1}$, being one of the free parameters in our work.

The rate of gas infall is assumed to be exponentially decreasing with
time:
\begin{equation}\label{eq:inf}
    f_{inf}(t)=A {\rm e}^{-t/\tau},
\end{equation}
where $A$ is the normalization constant which is constrained by the
boundary condition $\int^{t_G}_0 A{\rm e}^{-t/\tau}=1$, and $\tau$
is the infall timescale. So we can easily obtain the accretion rate
of an element $i$ through the formula
\begin{equation}\label{eq:inf2}
    \dot{G}_{i,inf}(t)=X_{i,inf}f_{inf}=\frac{X_{i,inf}{\rm e}
^{-t/\tau}}{\tau(1-{\rm e}^{-t_G/\tau})},
\end{equation}
$X_{i,inf}=0 (i\neq{\rm H, He})$ if primordial gas is assumed.

In our model, the galactic wind develops when the thermal energy of
the gas $E_g^{th}(t)$ exceeds its binding energy $E_g^b(t)$:
\begin{equation}\label{eq:EthEb}
    E_g^{th}(t)\geq E_g^b(t).
\end{equation}
The thermal energy of the gas is produced by SN explosions (both
Type II and Type Ia) and stellar winds:
\begin{equation}\label{eq:Eth}
E_g^{th}(t)=E_{SNII}^{th}(t)+E_{SNIa}^{th}(t)+E_{sw}^{th}(t).
\end{equation}
However, not all the energy produced in the above mentioned events
is stored into the ISM, since a fraction of it is lost by cooling.
In \cite{Bradamante98}'s work, the efficiencies of energy transfer
from SN and stellar winds into the ISM are the same,
$\eta_{SNII}=\eta_{SNIa}=\eta_{sw}=0.03$ (see their work for more
details). However, more recently, \cite{Recchi01} and
\cite{Recchi02} have shown that since SN Ia explosions occur in a
hotter and more rarefied medium, their energy can be more
efficiently thermalized into the ISM and, consequently, their
efficiency of energy transfer is higher. Therefore, in this work we
assume that the efficiencies of energy transfer are
$\eta_{SNII}=0.03, \eta_{SNIa}=0.8$ and $\eta_{sw}=0.03$ for SN II,
SN Ia and stellar winds respectively.

\begin{figure}[!t]
  \centering
   \includegraphics[width=0.45\textwidth]{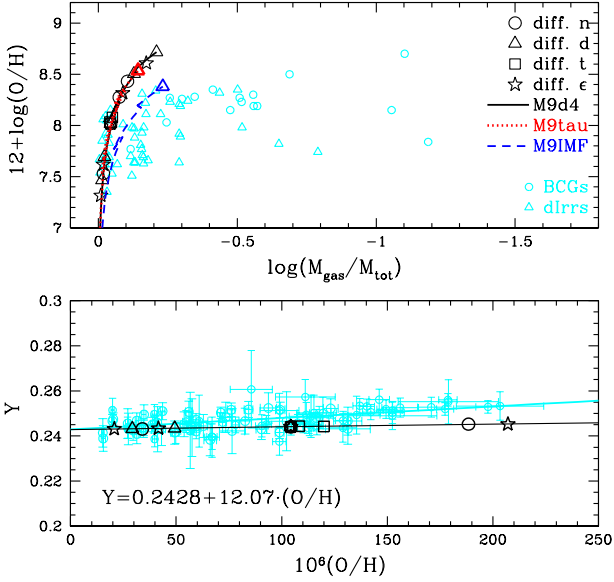}\\
  \caption{Present oxygen abundance vs. gas fraction ({\it upper}
panel) and $Y$ vs. oxygen abundance ({\it lower} panel) as predicted
by models without wind.
   All these models assume the same total infall mass $M_{inf}=10^9$\ms.
  {\it Big black open circles}: models with different numbers of bursts;
  {\it big black open triangles}: models with different durations of bursts;
  {\it big black open squares}: models with different occurrence times of bursts;
  {\it big black open pentacles}: models with different star formation efficiency.
  {\it Black solid, red dotted} and {\it blue-dash lines} in upper panel
are the evolutionary tracks of model M9d4, M9tau and M9IMF. The {\it
black-solid line} in lower panel is the best fit of all the model
points.
  The observational data are the same as in Fig.~\ref{Fig:obsmuZ} and
Fig.~\ref{Fig:obsYZ}, {\it cyan open circles} and {\it cyan open
triangles} represent BCDs and dIrrs respectively. The {\it cyan
solid line} is the best fit of the observational data.}
\label{Fig:nowdmuY}
\end{figure}

\begin{figure}[!t]
  \centering
   \includegraphics[width=0.45\textwidth]{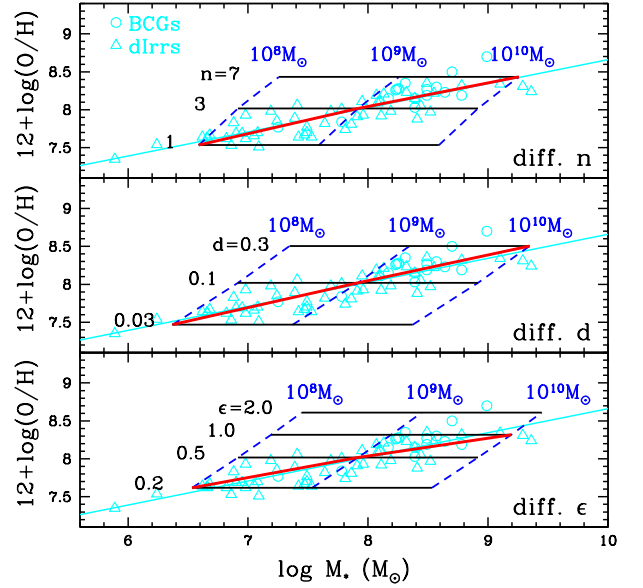}\\
  \caption{Present day oxygen abundance vs. $M_*$ as predicted by models without wind.
The upper (middle/lower) panel shows the model results with
different numbers of bursts (durations/SFE), along the {\it
blue-dashed lines} the number of bursts (durations/SFE) increases
from 2 to 7 (0.02 to 0.3/ 0.1 to 2), and along the {\it black-solid
line} the total infall mass $M_{inf}$ increases from $10^8$ to
$10^{10}$\ms. The {\it  red-thick-solid line} connects the models
which are consistent with the observational data, larger $n$
($d$/$\epsilon$) for more massive galaxy. The observational data are
the same as in Fig.~\ref{Fig:obsMsZ}, {\it cyan open circles} for
BCDs and {\it cyan open triangles} for dIrrs.}
  \label{Fig:nowdMsZ}
\end{figure}

To compute $ E_g^b(t)$, the binding energy of gas, we also followed
\cite{Bradamante98} and assumed that each galaxy has a dark matter
halo. The binding energy of gas is described as:
\begin{equation}
E_{Bgas}(t)=W_L(t)+W_{LD}(t)
\end{equation}
with:
\begin{equation}
W_L(t)=-0.5~G{ M_{gas}(t) M_L(t) \over r_L}
\end{equation}
which is the potential well due to the luminous matter and with:
\begin{equation}
W_{LD}(t)= -Gw_{LD}{M_{gas}(t) M_{dark} \over r_L}
\end{equation}
which represents the potential well due to the interaction between
dark and luminous matter, where $w_{LD} \sim {1 \over 2\pi}
S(1+1.37S)$, with $S= r_L/r_{D}$, being the ratio between the galaxy
effective radius ($r_L$) and the radius of the dark matter core
($r_D$) (see \citealt{Bertin92}).We assumed as in
\cite{Bradamante98} that the dark matter halo is 10 times more
massive than the luminous matter and that $S=0.3$.

The rate of gas loss via galactic wind for each element is assumed
to be simply proportional to the amount of gas present at the time
$t$:
\begin{equation}\label{eq:outf}
    \dot{G}_{i,out}(t)=w_i\lambda G(t)X_{i,out}(t),
\end{equation}
where $X_{i,out}(t)$, the abundance of the element $i$ in the wind,
is assumed to be same with $X_i(t)$, the abundance in the ISM;
$\lambda$ describes the efficiency of the galactic wind and has the
same units as $\epsilon$ (Gyr$^{-1}$); $w_i$ is the efficiency
weight of each element, hence $w_i\lambda$ is the effective wind
efficiency of the element $i$. $\lambda$ and $w_i$ are the other two
free parameters in our model. In this work, we have studied two
kinds of wind, the normal wind and the metal-enhanced wind. In the
case of the normal wind, all elements are lost in the same way,
i.e., $w_i=1$ for all elements, and we use $\lambda_w$ to denote the
wind efficiency in this case; however, wind provoked by SN explosion
could carry out more metals than H and He ($w_i>w_{\textrm{H,He}},
i\neq$ H,He), which is the so-called ``metal-enhanced'' wind
\citep{Mac99, Recchi01, Fujita03, Recchi08}, and we use
$\lambda_{mw}$ to denote the wind efficiency in this case.

\linespread{1.1}
\begin{table*}[!tb]
\centering
  \caption{Parameters of models without outflow}\label{Tab:nowd}
\begin{tabular}{llllllll}
\hline \hline
         &\multicolumn{3}{c}{Model name}        & SFE     & $n$ &$t$\tablefootmark{1} & $d$ \\
         \cline{2-4}
         & $10^8$\ms & $10^9$\ms & $10^{10}$\ms & Gyr$^{-1}$ && Gyr & Gyr \\
\hline
 diff. n &  M8n1 & M9n1 & M10n1 & 0.5   & 1 & 13               & 0.1*1  \\
         &  M8n2 & M9n2 & M10n2 & 0.5   & 3 & 1/7/13           & 0.1*3  \\
         &  M8n3 & M9n3 & M10n3 & 0.5   & 5 & 1/4/7/10/13     & 0.1*5  \\
         &  M8n4 & M9n4 & M10n4 & 0.5   & 7 & 1/3/5/7/9/11/13    & 0.1*7  \\
\hline
 diff. d &  M8d1 & M9d1 & M10d1 & 0.5   & 3 & 1/7/13          & 0.03*3  \\
         &  M8d2 & M9d2 & M10d2 & 0.5   & 3 & 1/7/13          & 0.1*3  \\
         &  M8d3 & M9d3 & M10d3 & 0.5   & 3 & 1/7/13          & 0.3*3  \\
         &       & M9d4 &       & 0.01  & 1 & 6.5             & 13  \\
\hline
 diff. t &  M8t1 & M9t1 & M10t1 & 0.5   & 3 & 1/7/13            & 0.1*3  \\
         &  M8t2 & M9t2 & M10t2 & 0.5   & 3 & 1/3/12           & 0.1*3  \\
         &  M8t3 & M9t3 & M10t3 & 0.5   & 3 & 8/10/13           & 0.1*3  \\
\hline
 diff. $\epsilon$&M8e1&M9e1&M10e1&0.1   & 3 & 1/7/13           & 0.1*3  \\
         &  M8e2 & M9e2 & M10e2 & 0.2   & 3 & 1/7/13           & 0.1*3  \\
         &  M8e3 & M9e3 & M10e3 & 1.0   & 3 & 1/7/13          & 0.1*3  \\
         &  M8e4 & M9e4 & M10e4 & 2.0   & 3 & 1/7/13         & 0.1*3  \\
\hline
 $\tau=10$ Gyr & & M9tau&       & 0.5   & 3 & 1/7/13           & 0.5*3  \\
 IMF$_{scalo86}$&& M9IMF&       & 0.5   & 3 & 1/7/1            & 0.5*3  \\
\hline \hline
\end{tabular} \\
\tablefoot{ \tablefoottext{1} The middle time of the burst.}
\end{table*}
\linespread{1.1}

\begin{figure*}[!t]
  \centering
   \includegraphics[width=0.58\textwidth]{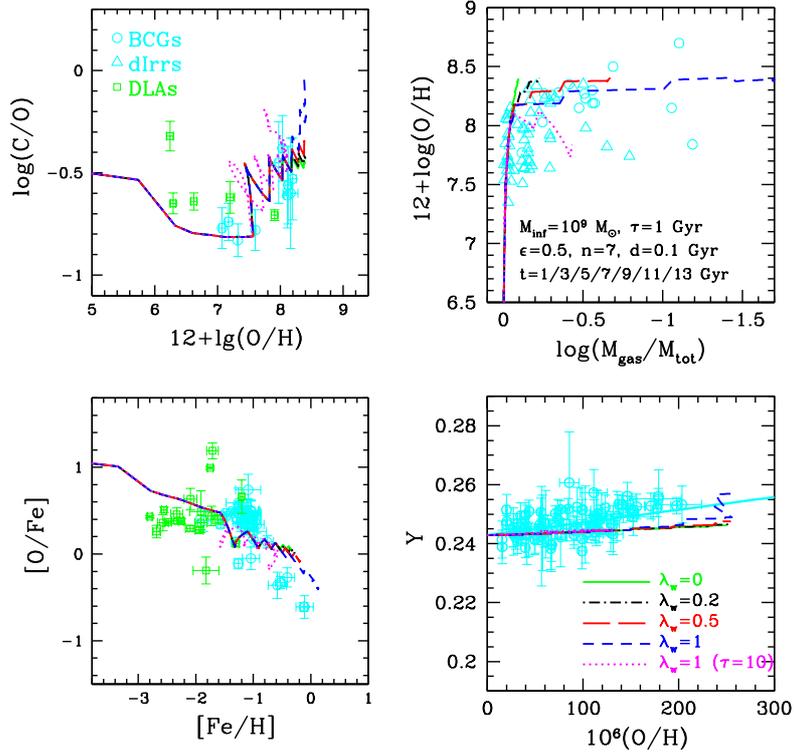}\\
  \caption{The evolutionary track as predicted by models with normal
wind. The left two panels are log(C/O) vs. 12+log(O/H) and [O/Fe]
vs. [Fe/H]; the right two panels are oxygen abundance vs. gas
fraction and $Y$ vs. (O/H). All the 5 models have same bursts
sequence ($\tau=1$ Gyr, $t=1/3/5/7/9/11/13$ Gyr, and $d=0.1$ Gyr for
each burst), but different wind efficiencies,
  {\it green-solid lines} for $\lambda_m=0$,
  {\it black-dash-dot lines} for $\lambda_m=0.2$,
  {\it red-long-dash lines} for $\lambda_m=0.5$,
  and {\it blue-short-dash lines} for $\lambda_m=1$.
A model with long infall timescale ($\tau=10$ Gyr, $\lambda_m=1$) is
shown in {\it magenta dotted line}.
   All these models assume the same total infall mass $M_{inf}=10^9$\ms.
  The observational data are the same as in  Fig.~\ref{Fig:obsmuZ},
Fig.~\ref{Fig:obsabund} and Fig.~\ref{Fig:obsYZ}, BCDs, dIrrs, and
DLAs are plotted in {\it cyan open circles, cyan open triangles} and
{\it green open squares} here.}
  \label{Fig:wdZmuY}
\end{figure*}

The initial mass function (IMF) is usually assumed to be constant
both in space and time in different galaxies, and can be expressed
as a power law of stellar mass as suggested first by
\cite{Salpeter55}:

\begin{equation}\label{eq:imf}
    \phi(m)=\phi_0 m^{-(1+x)},
\end{equation}
where $x=1.35$ for Salpeter IMF and $\phi_0$ is the normalization
constant which can be obtained by satisfying $\int m\phi(m)dm=1$ in
the mass range $0.1-100~M_{\odot}$. However, we tested also the
\cite{Scalo86} IMF:
\begin{equation} \label{eq:IMFscalo}
    \phi_{scalo86}(m)\propto\left\{
      \begin{aligned}
         & m^{-2.35}, \quad &(&0.1\leqslant m < 2) \\
         & m^{-2.7},  \quad &(&2\leqslant m \leqslant 100).
      \end{aligned}
    \right.
\end{equation}
which is a two-slope IMF and is steeper at the high mass end.

Stellar yields of different elements are important ingredients of
chemical evolution studies. In this work, we adopt stellar yields of
\cite{Woosley95} for massive stars and \cite{vanHoek97} for low- and
intermediate-mass stars. Both of them are metallicity-dependent.

\section{Model Results}

In order to understand the observed global properties and abundance
patterns of late-type dwarf galaxies, we have calculated several
models. The typical galaxy is assumed to be forming by continuous
infall of primordial gas and bursting star formation, as suggested
by several previous works (e.g., \citealt{Searle73, Matteucci83,
Marconi94, Bradamante98, Lanfranchi03, Romano06}). We also checked
the case of continuous star formation for these galaxies (see Sect.
4.5).

\subsection {Model without outflow}

Here we examine the models without outflow. Different numbers $n$,
durations $d$, times of the occurrence of bursts $t$, and different
star formation efficiencies $\epsilon$ have been tested. The
parameters adopted in the models are listed in Table~\ref{Tab:nowd}.
All the models assume short infall timescales ($\tau=1$ Gyr) except
model M9tau, and the galactic lifetime is taken to be 13 Gyr for all
the models. From the second to the forth columns there are the model
names, classified by different total infalling mass, from $10^8$\ms~
to $10^{10}$\ms~; the fifth column shows the SFE; the sixth column
the number of the bursts; the seventh column the middle time of each
burst; the eighth column the duration of each burst, where `` 0.1*3
'' means the duration of three bursts are the same, namely 0.1 Gyr.

In Fig.~\ref{Fig:nowdOFeH}, we show the evolutionary tracks of
[O/Fe] vs. [Fe/H] as predicted by models with a different number of
bursts(panel a), different times for the occurrence of the
bursts(panel b), different burst durations (panel c), and also
different SFEs (panel d). The most distinctive feature of the
bursting star formation scenario is the ``saw-tooth'' behaviour of
the tracks, which is caused by the different origins of
$\alpha$-elements and iron-peak elements. Oxygen is mainly
synthesized by massive stars, therefore its abundance increases only
during the bursting time, whereas iron is mainly synthesized by the
Type Ia supernovae and its abundance still increases after the burst
is over owing to the time delay of SN Ia explosions.

As we can see from Fig.~\ref{Fig:nowdOFeH}, different durations of
bursts and different SFEs among galaxies could be the explanation of
the scatter in the data. In panel c, as a comparison, we also plot
the model results for a low continuous star formation process (model
M9d4). Clearly in this case the saw-tooth behaviour disappears.

If we compare the present oxygen abundance vs. the gas fraction as
predicted by the above models with the observations
(Fig.~\ref{Fig:nowdmuY}, upper panel), we can see that no matter how
the star formation history changes (different $n, t, d, \epsilon$),
the model results always stay along the same curve. As first
suggested by \cite{Matteucci83}, in order to explain the spread in
this diagram one should necessarily claim a variation of the IMF, or
of the wind rate, or of the infall rate.
Therefore, we examined the other IMF \citep{Scalo86} which is
steeper than Salpeter IMF at the massive end.
However, the model results in Fig.~\ref{Fig:nowdmuY} still cannot
explain the observed lowest oxygen abundance at the same $\mu$ or
the lowest $\mu$ at the same oxygen abundance. Therefore, unless one
assumes unrealistically steeper IMFs, the observed $\mu-Z$ strongly
implies that there should have other mechanisms operating in the
galaxy which can reduce the O abundance or the gas fraction.

In the lower panel of Fig.~\ref{Fig:nowdmuY}, it is shown the
$Y-$(O/H) relation as predicted by our model  which is consistent
with the observations, although a little flatter than the observed
best fit. The best fit to these model results is
$Y=0.2428+12.07$(O/H).

\begin{figure}[!t]
  \centering
   \includegraphics[width=0.45\textwidth]{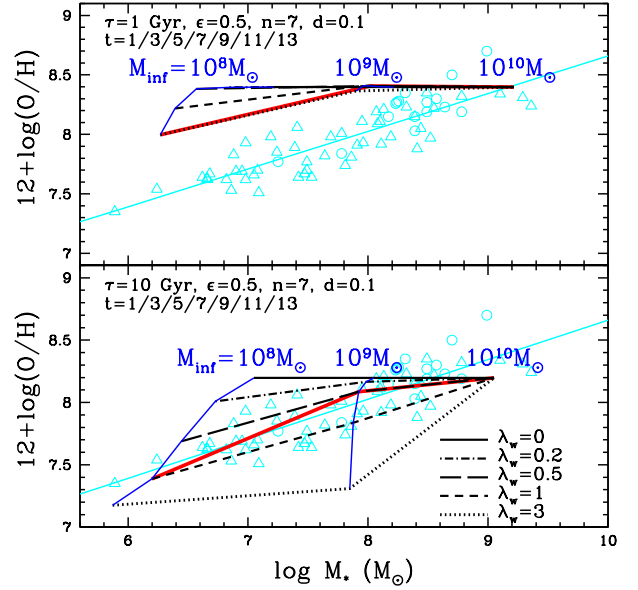}\\
  \caption{Present day oxygen abundance vs. $M_*$ as predicted by
models with normal wind. All the models contain 7 bursts and the
same burst sequence ($\epsilon=0.2, n=7, t=1/3/5/7/9/11/13, d=0.1$
for each burst), and the total infall mass varies from
$M_{inf}=10^8$ to $10^{10}$\ms. The upper panel shows results with
short infall timescale ($\tau=1$ Gyr), whereas the lower one shows
results with long timescale ($\tau=10$ Gyr). The predicted $M-Z$
relation with different strength of normal wind are shown in black
lines: {\it black solid, dash-dot, long-dash, short-dash, and dotted
lines} are for $\lambda_m=0, 0.2, 0.5, 1, 3$ respectively. Models
with same $M_{inf}$ are connected by {\it blue lines}. An increasing
wind efficiency to less massive galaxies is shown in {\it red-solid
line}. Observational data are shown in {\it cyan open circles}
(BCDs) and {\it cyan open triangles} (dIrrs), the {\it cyan solid
line} is the best fit of all data, same as in
Fig.~\ref{Fig:obsMsZ}.}
  \label{Fig:wdMsZ}
\end{figure}

\begin{figure*}[!t]
  \centering
   \includegraphics[width=0.7\textwidth]{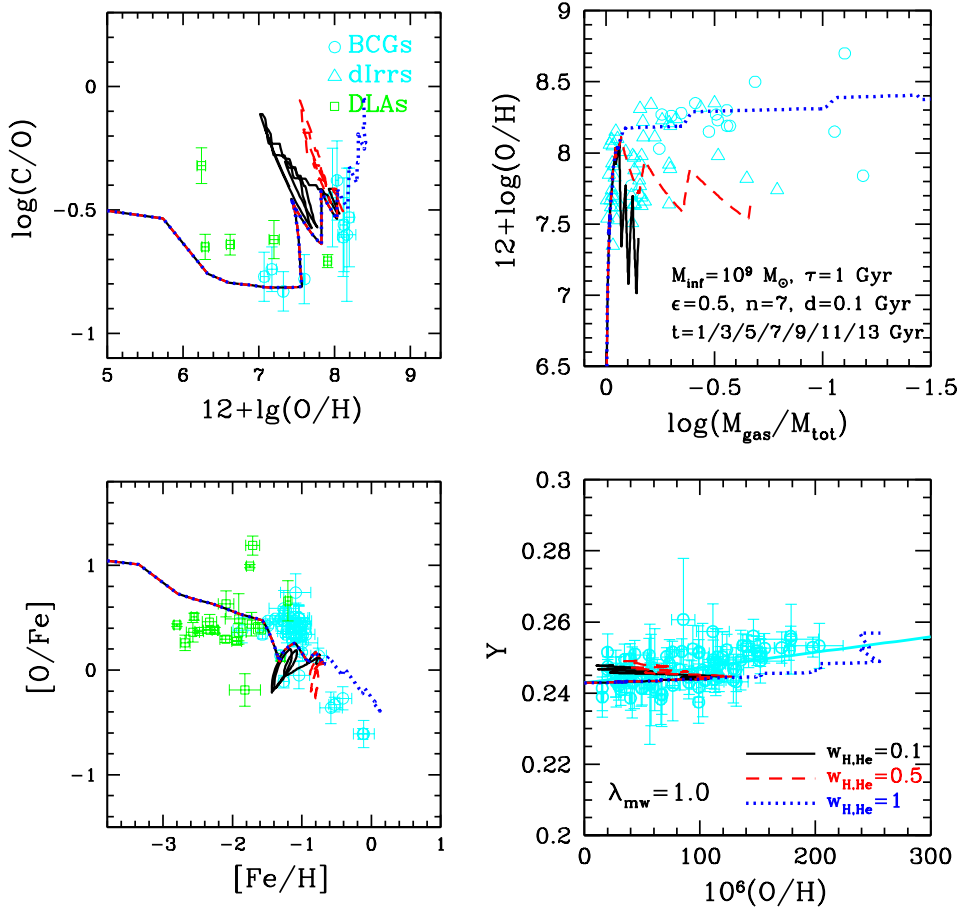}\\
  \caption{The evolutionary track as predicted by models with
various amounts of metal enhancements. The left two panels are
log(C/O) vs. 12+log(O/H) and [O/Fe] vs. [Fe/H]; the right two panels
are oxygen abundance vs. gas fraction and $Y$ vs. (O/H). All the 3
models have same bursts sequence ($t=1/3/5/7/9/11/13$ Gyr, and
$d=0.1$ Gyr for each burst), but different degrees of metal
enhancement. The {\it black-solid lines, red-dash lines, blue-dot
lines} represent $w_{\rm H, He}=0.1, 0.5, 1$ respectively.
  All these models assume the same total infall mass $M_{inf}=10^9$\ms.
The observational data are the same as in Fig.~\ref{Fig:obsmuZ},
Fig.~\ref{Fig:obsabund} and Fig.~\ref{Fig:obsYZ}, BCDs, dIrrs, and
DLAs are plotted in {\it cyan open circles, cyan open triangles} and
{\it green open squares} here.}
  \label{Fig:mwdZmuY1}
\end{figure*}

\begin{figure*}[!t]
  \centering
   \includegraphics[width=0.7\textwidth]{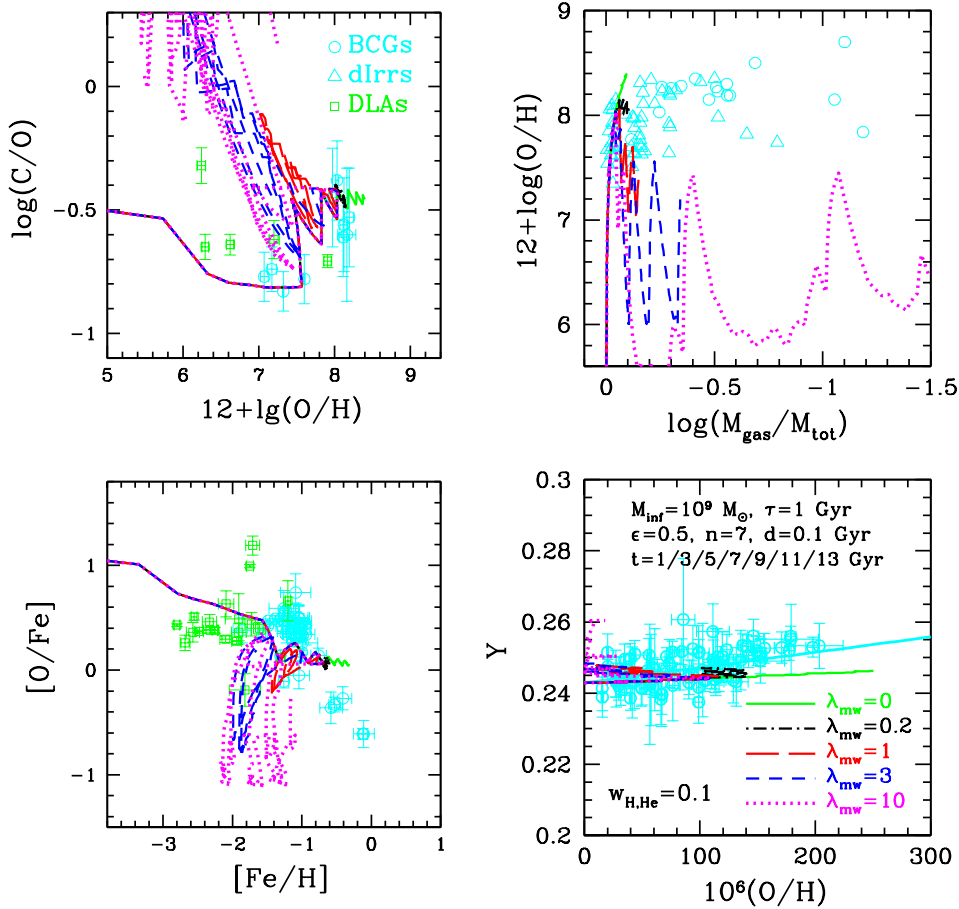}\\
  \caption{The evolutionary track as predicted by models with
various metal-enhanced wind efficiencies. The left two panels are
log(C/O) vs. 12+log(O/H) and [O/Fe] vs. [Fe/H]; the right two panels
are oxygen abundance vs. gas fraction and $Y$ vs. (O/H). All the 5
models have same bursts sequence ($t=1/3/5/7/9/11/13$ Gyr, and
$d=0.1$ Gyr for each burst), but different wind efficiencies.
  {\it Green-solid lines}: $\lambda_{mw}=0$;
  {\it black- dash-dot lines}: $\lambda_{mw}=0.2$;
  {\it red-long-dash lines}: $\lambda_{mw}=1$;
  {\it blue-short-dash lines}: $\lambda_{mw}=3$;
  {\it magenta-dot lines}: $\lambda_{mw}=10$.
  All these models assume the same total infall mass $M_{inf}=10^9$\ms.
  The observational data are the same as in Fig.~\ref{Fig:obsmuZ},
Fig.~\ref{Fig:obsabund} and Fig.~\ref{Fig:obsYZ}, BCDs, dIrrs, and
DLAs are plotted in {\it cyan open circles, cyan open triangles} and
{\it green open squares} here.}
  \label{Fig:mwdZmuY2}
\end{figure*}

Since we assume a linear correlation between star formation rate and
gas, the models with different total infalling masses are
self-similar. Therefore, when the abundance ratios and gas fraction
are examined (Figs.~\ref{Fig:nowdOFeH} and \ref{Fig:nowdmuY}), the
three series of models (M8, M9, and M10) are overlapping.

In Fig.~\ref{Fig:nowdMsZ} we plot the mass-metallicity relations
predicted by models without wind. We run models for three different
infall masses ($10^{8}$, $10^{9}$, $10^{10}$\ms). The effects of
different numbers of bursts ($n=1, 3, 7$), different durations
($d=0.03, 0.1, 0.3$) and different SFEs ($\epsilon=0.2, 0.5, 1.0,
2.0$) as functions of galactic mass are shown. It is evident from
Fig.~\ref{Fig:nowdMsZ} that our models can very well reproduce the
M-Z relation even without galactic wind but just assuming an
increase of the number, or duration of bursts, or the efficiency of
SF.

\subsection {Model with normal wind}

If the galactic wind  has the same chemical composition as the
well-mixed ISM, i.e. $w_i=1$ for all the elements, we call it
``normal wind''.

In Fig.~\ref{Fig:wdZmuY}, we show the the evolutionary tracks
predicted by models with normal wind; abundance ratios of log(C/O)
vs. 12+log(O/H) and [O/Fe] vs. [Fe/H] are on the left side, while
$\mu-Z$ and $Y-Z$ relations on the right side. The models have the
same total infall mass ($M_{inf}=10^9$\ms) and same bursts sequence
($t=1/3/5/7/9/11/13$ Gyr, with $d=0.1$ Gyr for each burst), but
different wind efficiencies ($\lambda_w=0, 0.2, 0.5, 1.0$). Oxygen
is produced by massive stars, therefore no oxygen will be ejected
into the ISM after star formation ceases. On the other hand,
elements, such as C and N produced by low- and intermediate-mass
stars, and Fe mainly produced by SN Ia explosion, are continuously
polluting the ISM after the star formation stops, owing to their
long lifetime. Therefore, the decrease of the mass of gas (i.e., H
and He) and the $\alpha$-elements lost with the wind will result in
a dramatic increasing of the abundance of the ``time-delayed''
elements. The stronger the wind, the higher the abundance of C or Fe
relative to O predicted by the models.

The main effect of normal winds is to decrease the gas fraction with
a smaller effect on  the O/H abundance, as we can see from the
12+log(O/H)-$\mu$ relation (upper right panel of
Fig.~\ref{Fig:wdZmuY}). This means that models with normal wind
cannot explain the whole spread in O/H observed at a given $\mu$ for
these galaxies. There are two possible reasons for that. One is the
same wind efficiency (i.e., $w_i\lambda_w$) for both oxygen and
hydrogen in the normal wind, so that both O and H decrease at the
same time. The other one is the short infall time scale assumed
($\tau=1$ Gyr). In this case, no primordial gas falls into the
galaxies to dilute the ISM at late evolutionary times.  Therefore,
we also developed a model with long infall time scale ($\tau=10$
Gyr, magenta dotted lines in Fig.~\ref{Fig:wdZmuY}). It is clear
that in this model the metallicity decreases in the interburst time.
Actually, the infall of primordial gas (i.e., H and He) results in a
lower mass loss rate of H and He than metals, similar to the
metal-enhanced wind case which will be further discussed in the next
section.  A very strong normal wind (e.g. $\lambda_w>0.5$) seems
unlikely in late-type dwarf galaxies since it would lose a large
amount of gas, and hence it would predict a too low gas fraction, as
it is evident in Fig.~\ref{Fig:wdZmuY}. In the lower right panel of
Fig.~\ref{Fig:wdZmuY}, the predicted Y vs. (O/H) relation is shown
and it is consistent with the observational data at the low
metallicity, because the wind does not develop yet when the galaxy
is still very metal poor. However, after the wind, an increase of
the helium abundance as well as of the abundances of elements
produced on long timescale occurs, especially in the case of a
strong wind which produces a very small final gas fraction.

In the last section, we have pointed out that the observed $M-Z$
relation could indicate more star bursts or longer duration of each
burst or higher SFE in more massive galaxies, if no outflow takes
place. Now we examine the possibility of the normal wind being the
explanation of the $M-Z$ relation, as suggested by many previous
authors. In Fig.~\ref{Fig:wdMsZ}, we take the models with 7 bursts
for example (i.e., Model M8n4, M9n4, M10n4 in no-wind case) but
different strengthes of normal wind are introduced ($\lambda_{w}=0,
0.2, 0.5, 1, 3$). It is evident from the upper panel of
Fig.~\ref{Fig:wdMsZ} that by varying only the efficiency of a normal
wind one cannot reproduce the $M-Z$ relation, unless other
parameters, such as the efficiency of SF or the number of bursts,
are assumed to vary as functions of the galactic mass. When a long
infall timescale is adopted ($\tau=10$ Gyr, lower panel of
Fig.~\ref{Fig:wdMsZ}), less stars are formed in each model due to
the slow gas accretion process, and the wind could not be induced in
the high mass systems ($M_{inf}\approx10^{10}$\ms). However in the
low mass galaxy ($M_{inf}=10^8$\ms), where the wind could develop,
the newly infalling gas dilutes the ISM effectively. Therefore, the
general trend of the $M-Z$ relation could be reproduced by combining
the normal wind with a slow accretion process.

In summary, the normal wind can strongly reduce the gas fraction but
it cannot reduce sensibly the O/H. To explain the spread observed in
O/H at the same $\mu$ we should invoke other mechanisms, such as a
continuous supplement of primordial gas or a metal-enhanced wind
(see Sect. 4.3), both of which imply a lower mass loss rate for H
and He relative to metals.

\subsection {Model with metal-enhanced wind}

The galactic wind, mainly induced by SN explosion, could blow
preferentially the metal-enriched gas out of the galaxy, which means
metals are lost more efficiently than the gas (H and He). We define
the wind ``metal-enhanced'' when the abundances of metals it carries
out are higher than in the ISM. Metal-enhanced winds have been
already suggested by several dynamical works (e.g. \citealt{Mac99,
Recchi01, Recchi02}). In our models we simply assume a higher wind
efficiency weight $w_i$ for heavy elements than H and He. In
particular, we adopt $w_i=1 (i\neq\rm H, He)$, $w_{\rm H,He}<1$.

\begin{figure}[!t]
  \centering
   \includegraphics[width=0.4\textwidth]{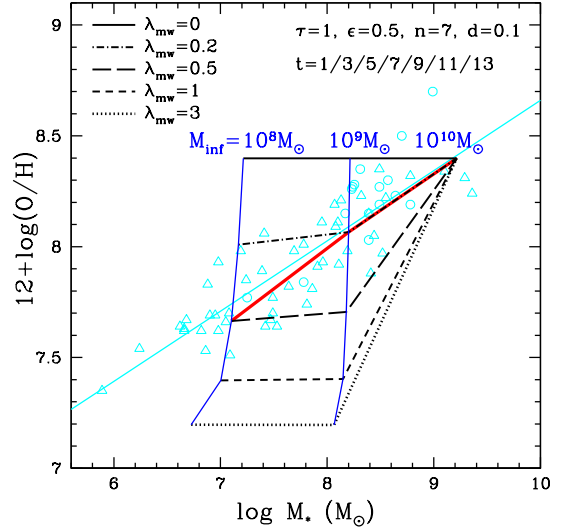}\\
  \caption{Present day oxygen abundance vs. $M_*$ as predicted by
models with metal-enhanced wind ($w_{\rm H,He}=0.1$). All the models
contain 7 bursts and the same star formation history ($\epsilon=0.2,
n=7, t=1/3/5/7/9/11/13, d=0.1$ for each burst), and the total infall
mass varies from $M_{inf}=10^8$ to $10^{10}$\ms. The predicted $M-Z$
relation with different strength of metal-enhanced wind are shown in
different lines: {\it black dash-dot, long-dash, short-dash, and
dotted lines} are for $\lambda_{mw}=0, 0.2, 1, 3$ respective. Models
with same $M_{inf}$ are connected by {\it blue lines}. An increasing
wind efficiency to less massive galaxies is shown in {\it red-solid
line} which can fit the data very well . Observational data are
shown in {\it cyan open circles} (BCDs) and {\it cyan open
triangles} (dIrrs), the {\it cyan solid line} is the best fit of all
data, same as in Fig.~\ref{Fig:obsMsZ}.}
  \label{Fig:mwdMsZ}
\end{figure}

The wind models with various amounts of metal enhancements are shown
in Fig.~\ref{Fig:mwdZmuY1}. All the models have same input
parameters ($\epsilon=0.5$, 7 bursts and the duration is 0.1 Gyr for
each one) except for $w_{\rm H,He}$. The model experiencing a highly
enriched wind ($w_{\rm H,He}=0.1$) loses very little gas. We show
also models with mild metal-enhanced wind ($w_{\rm H,He}=0.5$) and
normal wind ($w_{\rm H,He}=w_{\rm O}=1$). The evolutionary tracks
for the abundance ratios show a loop if the wind is metal-enriched.
When the wind starts, oxygen is lost more efficiently than hydrogen,
hence the oxygen abundance within the galaxy decreases with time in
the interburst phase, and elements such as C and Fe will show
increasing abundances relative to oxygen owing to their delayed
restoration into the ISM. This trend continues until the new burst
occurs. Because of the newly produced oxygen supplied to the ISM the
O abundance increases and, as a consequence,  the abundances of
other elements relative to oxygen decrease. Therefore, the
evolutionary track shows a loop. The lower the $w_{\rm H,He}$, the
more the metals lost, the lower the value that the O abundance
reaches.

The metal-enhanced wind has also a dramatic influence on the $\mu-Z$
relation, as we shown in the upper right panel of
Fig.~\ref{Fig:mwdZmuY1}. The normal wind mainly reduces the gas
fraction rather than the abundance, whereas the metal-enhanced wind
is very powerful in reducing the metallicity of the galaxy.


\linespread{1.3}
\begin{table*}[!t]
\centering
  \caption{Parameters of the best models ($w_{\rm H,He}=0.3$).}\label{Tab:best}
\begin{tabular}{ccccclllll}
\hline \hline
  \multicolumn{3}{c}{Model name}       & $\lambda_{mw}$ & SFE   & $n$ & $t$\tablefootmark{1} & \multicolumn{3}{c}{$d$\tablefootmark{2} (Gyr)} \\
  \cline{1-3}   \cline{8-10}
  $10^8$\ms & $10^9$\ms & $10^{10}$\ms & Gyr$^{-1}$ & Gyr$^{-1}$&   & Gyr             & $10^8$\ms & $10^9$\ms & $10^{10}$\ms  \\
\hline
  M8b1      & M9b1      & M10b1        & 0.8          & 0.5       & 3 & 3/9/13          & 0.1*3    & 0.3*3     & 0.9*3  \\
  M8b2      & M9b2      & M10b2        & 0.8          & 0.5       & 5 & 1/3/7/10/13     & 0.1*5    & 0.3*5     & 0.9*5  \\
  M8b3      & M9b3      & M10b3        & 0.8          & 0.5       & 7 & 1/3/5/7/9/11/13 & 0.1*7    & 0.3*6     & 0.9*7  \\
  M8b4      & M9b4      & M10b4        & 0.8          & 0.5       & 9 & 1/2.5/4/5.5/7/8.5/10/11.5/13 & 0.1*9 & 0.3*9 & 0.9*9 \\
\hline
            & DLA1      &              & 0.8          & 0.5       & 5 & 1/3/7/10/13     &           & 0.01/0.2*4&\\
            & DLA2      &              & 0.8          & 0.5       & 5 & 1/3/7/10/13     &           & 0.02/0.2*4&\\
            & DLA3      &              & 0.8          & 0.5       & 5 & 1/3/7/10/13     &           & 0.05/0.2*4&\\
            & DLA4      &              & 0.8          & 0.5       & 5 & 1/3/7/10/13     &           & 0.1/0.2*4&\\
\hline \hline
\end{tabular} \\
\tablefoot{ \tablefoottext{1} The middle time of the burst}
\end{table*}
\linespread{1.1}

\begin{figure*}[!t]
  \centering
   \includegraphics[width=0.8\textwidth]{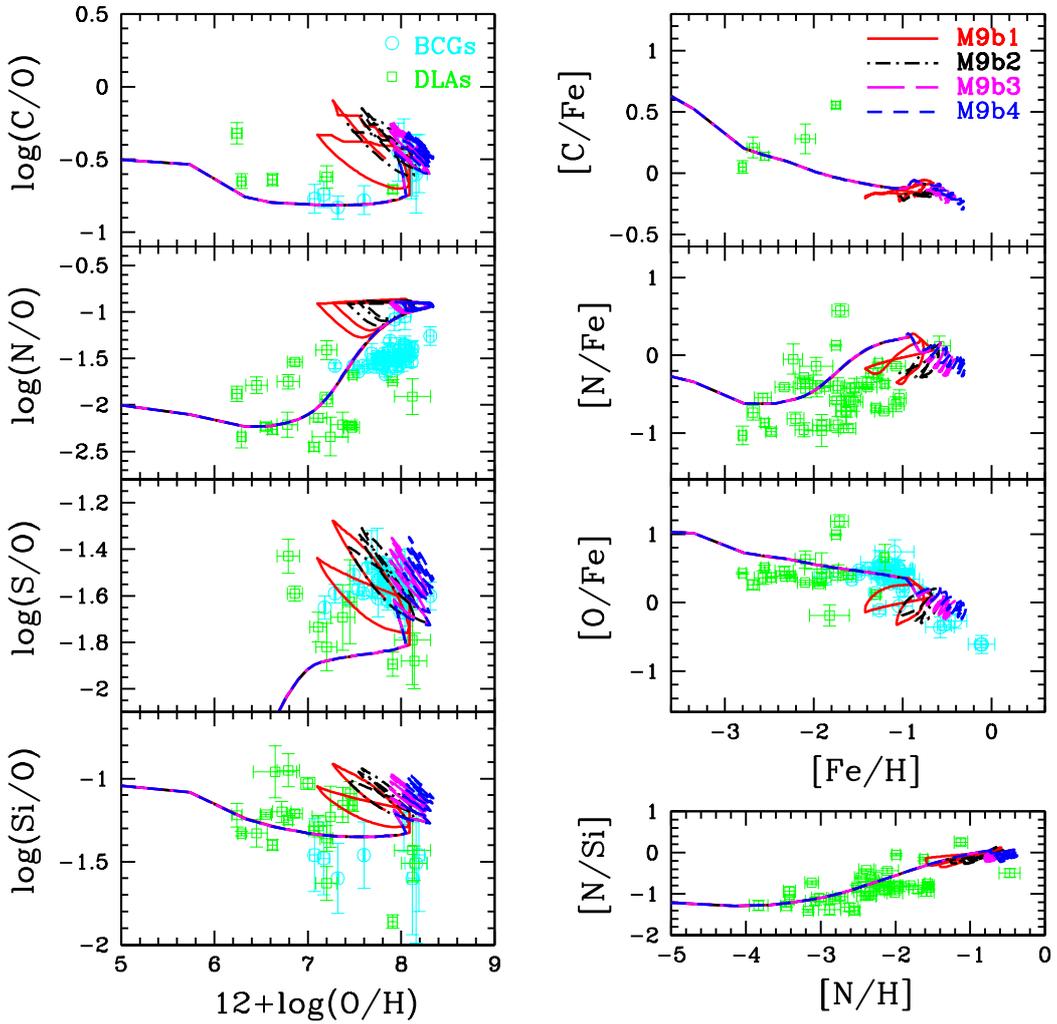}\\
  \caption{The evolutionary track of abundance ratios of C/O, N/O,
S/O, Si/O, C/Fe, N/Fe, O/Fe, and N/Si as predicted by our best
models ($M_{inf}=10^9$\ms~) with metal-enhanced wind ($w_{\rm
H,He}=0.3$). The {\it red-solid, black-dash-dot, magenta-long-dash }
and {\it blue-short-dash} lines are the results of M9b1, M9b2, M9b3
and M9b4 in Table.~\ref{Tab:best}. The observational data are same
as in Fig.~\ref{Fig:obsabund}, {\it cyan open circles} for BCDs and
{\it green open squares} for DLAs. }
  \label{Fig:bestabund}
\end{figure*}

In Fig.~\ref{Fig:mwdZmuY2}, models with different strengths
($\lambda_{mw}=0, 0.2, 1, 3, 10$) of highly enriched winds ($w_{\rm
H,He}=0.1$) are shown.  We plot the evolutionary tracks of models
with same bursting history as in Fig.~\ref{Fig:mwdZmuY1}. A stronger
wind not only reduces the abundances and increases the ratio between
the long recycling term elements and the short ones, but also
decreases the gas fraction dramatically. It is worth to point out
that although the wind efficiencies $\lambda_{mw}$ adopted here are
much higher than the ones of normal wind case $\lambda_{w}$, the gas
is lost less effectively. To compare the results of normal and
metal-enhanced wind models, one should assume a larger
$\lambda_{mw}$ for the latter case. For example, $\lambda_{mw}=10$
for metal-enhanced winds is then multiplied by $w_{\rm H,He}=0.1$,
therefore it is comparable with the case normal wind and
$\lambda_{w}=1$.

In the lower right panels of both Fig.~\ref{Fig:mwdZmuY1} and
Fig.~\ref{Fig:mwdZmuY2} we show the $Y-$(O/H) relations of galaxies
with different degrees of enriched winds $w_{\rm H,He}$ and
different wind efficiencies $\lambda_{mw}$, but the same formation
histories. By comparing with Fig.~\ref{Fig:wdZmuY}, the present-day
oxygen abundances are lower as we expect. In this scenario, a very
high helium abundance can be reached at low metallicity level,
especially when the wind is very strong, because most of it stays
inside the galaxy while heavy elements are lost. The present-day
$Y-$(O/H) relation predicted by these models do not stay on a
straight line in the low metellicity region, even if the unrealistic
models (very strong wind $\lambda_{mw}=10$ cases) are ruled out
considering their disagreement with other observational constraints.
Therefore, the observed scatter of $Y-$(O/H) relation may be caused
by metal-enhanced winds. Based on our model predictions, we suggest
to fit the lower envelop of the observational data, when one derives
the primordial helium $Y_p$, because it may not be affected by the
wind, hence the extrapolation to $Z=0$ will be more close to the
real $Y_p$.

\linespread{1.3}
\begin{table}[!t]
\centering
  \caption{ The maximum and present-day values of the SN Ia rates by
number as predicted by the best models.}\label{Tab:SNIa}
\begin{tabular}{cccc}
\hline \hline
  Model & number & \multicolumn{2}{c}{$r_{\rm SN Ia}$(century$^{-1}$)}        \\
  \cline{3-4}
  name  & of burst  & maximum & present day  \\
\hline
  M9b1  & 3 & 0.0218   & 0.0016  \\
  M9b2  & 5 & 0.0229   & 0.0012  \\
  M9b3  & 7 & 0.0229   & 0.0015  \\
  M9b4  & 9 & 0.0237   & 0.0016  \\
\hline \hline
\end{tabular} \\
\end{table}
\linespread{1.1}

\begin{figure}[!t]
  \centering
   \includegraphics[width=0.4\textwidth]{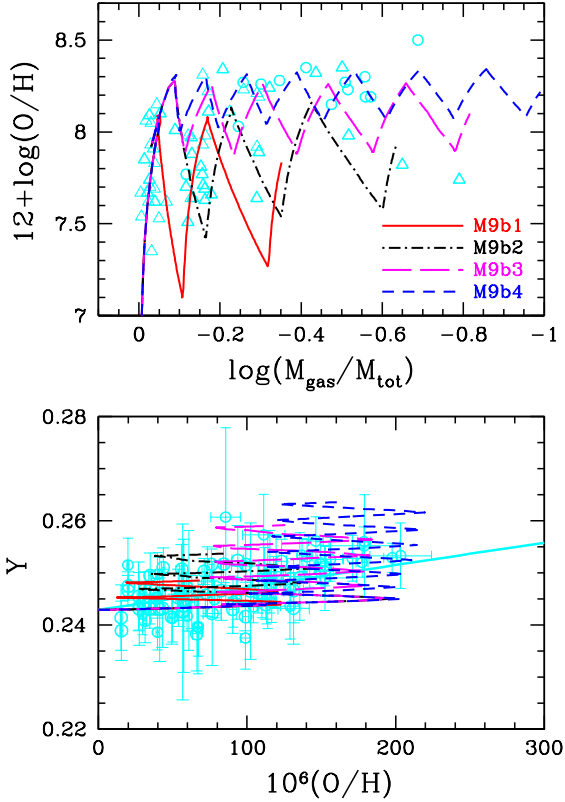}\\
  \caption{The evolutionary tracks of 12+log(O/H) vs. gas fraction
({\it upper} panel) and $Y-$(O/H) relation ({\it lower} panel) as
predicted by our best models ($M_{inf}=10^9$\ms~) with
metal-enhanced wind ($w_{\rm H,He}=0.3$). The {\it red-solid,
black-dash-dot, magenta-long-dash} and {\it blue-short-dash} lines
are the results of M9b1, M9b2, M9b3 and M9b4 in
Table.~\ref{Tab:best}. The observational data are same as in
Fig.~\ref{Fig:obsmuZ} and Fig.~\ref{Fig:obsYZ}, {\it cyan open
circles} for BCDs and {\it cyan open triangles} for dIrrs.}
  \label{Fig:bestmuY}
\end{figure}

\begin{figure}[!t]
  \centering
   \includegraphics[width=0.4\textwidth]{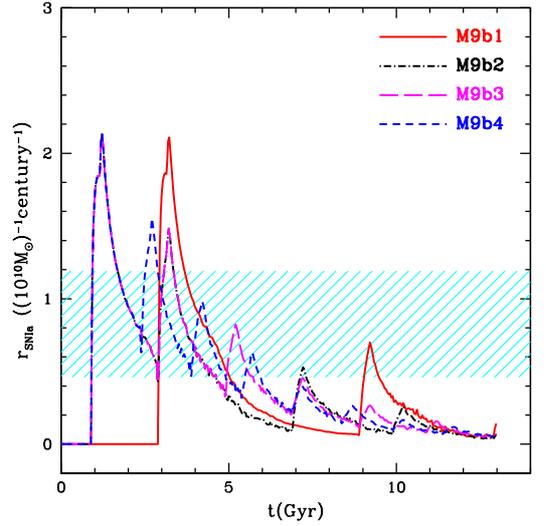}\\
  \caption{The SN Ia rate normalized to the galaxy stellar mass as
a function of time as predicted by our best models
($M_{inf}=10^9$\ms). The {\it red-solid, black-dash-dot,
magenta-long-dash} and {\it blue-short-dash} lines are for model
M9b1, M9b2, M9b3 and M9b4 respectively. As a comparison, the {\it
shaded area} shows the observed range of normalized SN Ia rate in
the Irr galaxies, $0.77^{+0.42}_{-0.31}$ per century and per
$10^{10}$\ms \citep{Mannucci05}.}
  \label{Fig:bestSNIa}
\end{figure}

\begin{figure}[!t]
  \centering
   \includegraphics[width=0.4\textwidth]{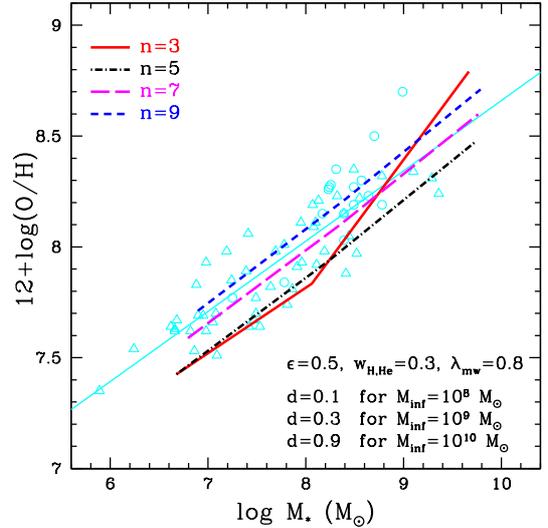}\\
  \caption{The mass-metallicity relation as predicted by our best models.
The total infalling masses range from $M_{inf}=10^8$ to
$10^{10}$\ms), and more massive one prefers longer duration of
burst. The results with different numbers of burst are shown in
different lines, {\it red-solid, black-dash-dot, magenta-long-dash }
and {\it blue-short-dash} lines are for $n=3, 5, 7, 9$ respectively.
The observational data of BCDs ({\it cyan open circles}) and dIrrs
({\it cyan open triangles}) are the same as in
Fig.~\ref{Fig:obsMsZ}.}
  \label{Fig:bestMsZ}
\end{figure}

Fig.~\ref{Fig:mwdMsZ} is the same as Fig.~\ref{Fig:wdMsZ} but for
models with metal-enhanced wind. Unlike in the normal wind case, the
metal-enhanced one is very effective in reducing the oxygen
abundance, hence in creating the $M-Z$ relation.
The stronger the wind
efficiency, the steeper the predicted $M-Z$. Therefore, models with
 an increasing wind efficiency to less massive galaxies are
consistent with the observations very well. As a conclusion, the
observed $M-Z$ relation could be caused by metal-enhanced winds with
mild strength (e.g., $\lambda_{mw}\le 1$).

In summary, metal-enhanced winds should take place in late-type
dwarf galaxies, and they play an important role in removing gas,
especially the metals, out of the galaxies.

\subsection{Best models}

\begin{figure*}[!t]
  \centering
   \includegraphics[width=0.8\textwidth]{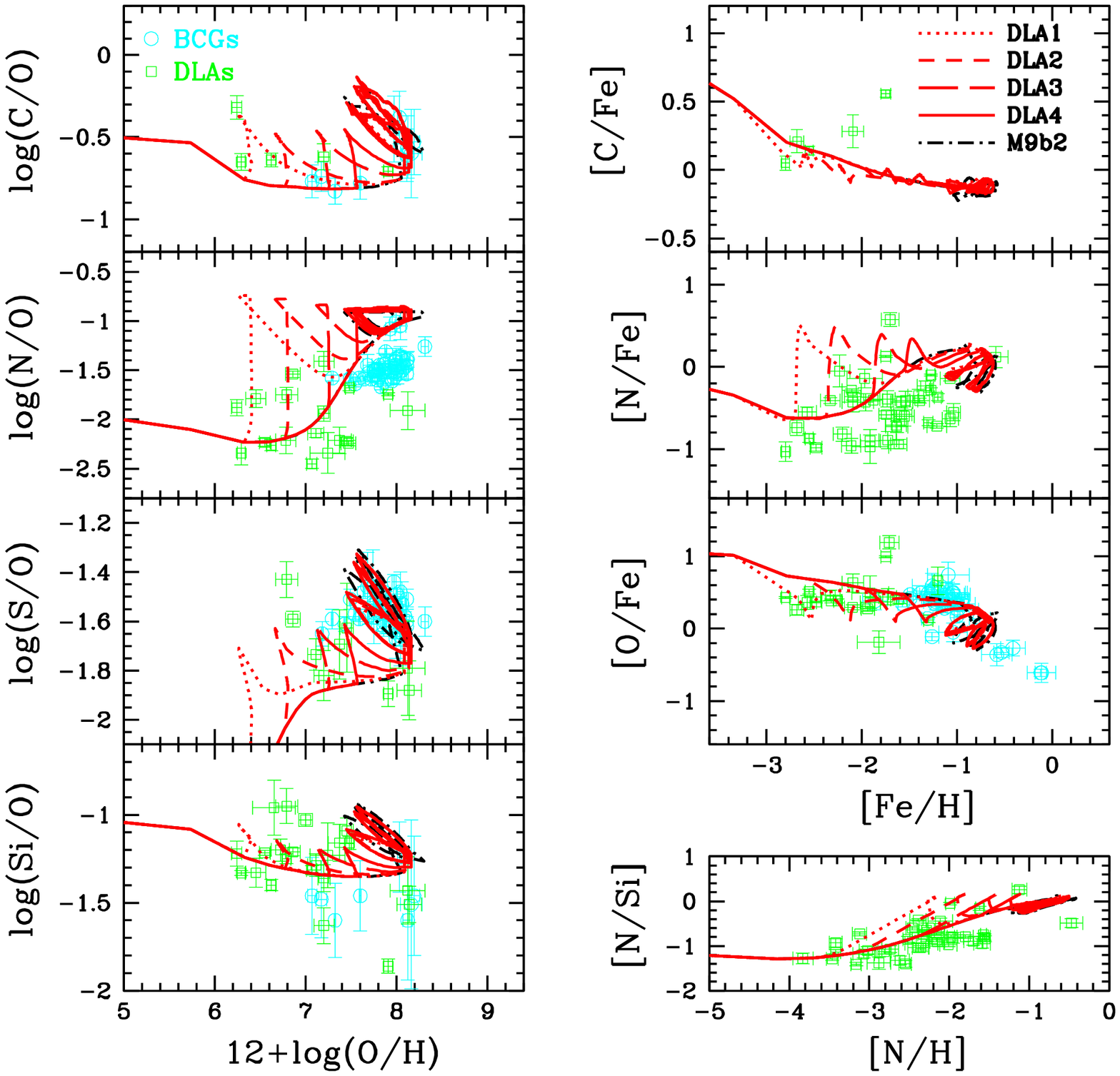}\\
  \caption{The evolutionary track of abundance ratios as predicted
by our DLA models. All of these models assume same SFHs as model
M9b3 except for a shorter duration of the first burst used. The red
{\it dotted, short-dash, long-dash} and {\it solid} lines are for
models DLA1 to DLA4 whose durations of the first burst are $d=0.01,
0.02, 0.05, 0.1$ Gyr respectively. As a comparison, we also show the
best model M9b2 of dwarf galaxy in {black-dash-dot} lines.
  All these models assume the same total infall mass $M_{inf}=10^9$\ms.
{\it Cyan open circles} are BCDs and {\it green open squares} are
DLAs, same as in Fig.~\ref{Fig:obsabund}. }
  \label{Fig:DLAabund}
\end{figure*}

\begin{figure*}[!t]
  \centering
   \includegraphics[width=0.8\textwidth]{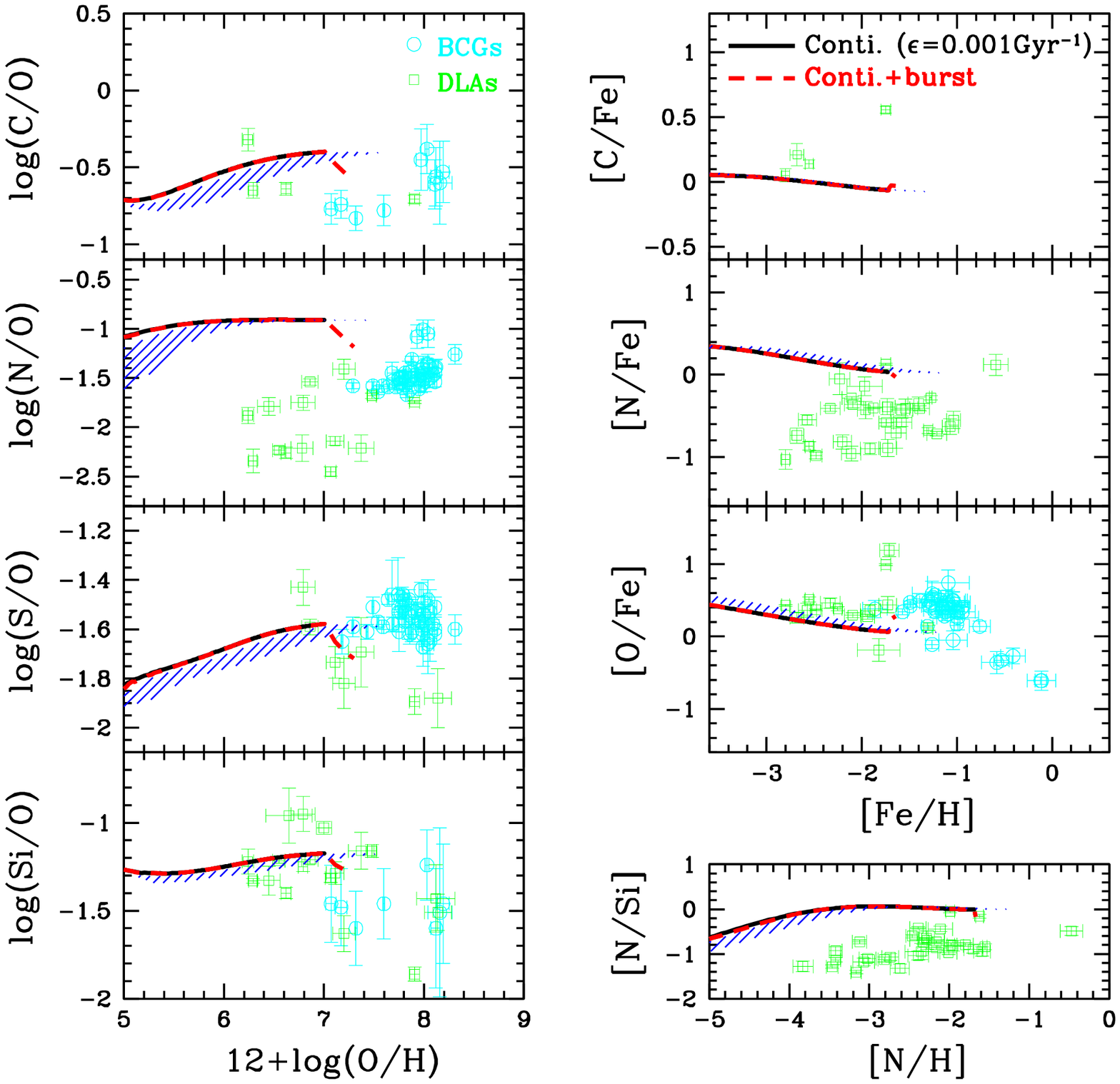}\\
  \caption{The evolutionary track of abundance ratios as predicted
by the models of \citet{Legrand00a}. A mild continuous SF with
$\epsilon=0.001$ Gyr$^{-1}$ is shown in {\it black-solid lines},
whereas a mild continuous SF ($\epsilon=0.001$ Gyr$^{-1}$) combined
with a current burst ($SFR=0.023$\ms yr$^{-1}$, i.e.,
$\epsilon=0.88$ Gyr$^{-1}$ if $M_{\rm HI}=2.6\times10^7$\ms, during
the last 20 Myrs) is shown in {\it red-dash lines}. The {\it blue
shade areas} show the continuous SF with
$1.45\times10^{-3}\le\epsilon\le3.85\times10^{-3}$ Gyr$^{-1}$ which
is derived from the observed SFR and $M_{\rm HI}$ data of IZw 18(see
Table~\ref{Tab:DLA} of \citealt{Legrand00a}). {\it Cyan open
circles} are BCDs and {\it green open squares} are DLAs, same as in
Fig.~\ref{Fig:obsabund}. }
  \label{Fig:Legrandabund}
\end{figure*}

\begin{figure*}[!t]
  \centering
   \includegraphics[width=0.8\textwidth]{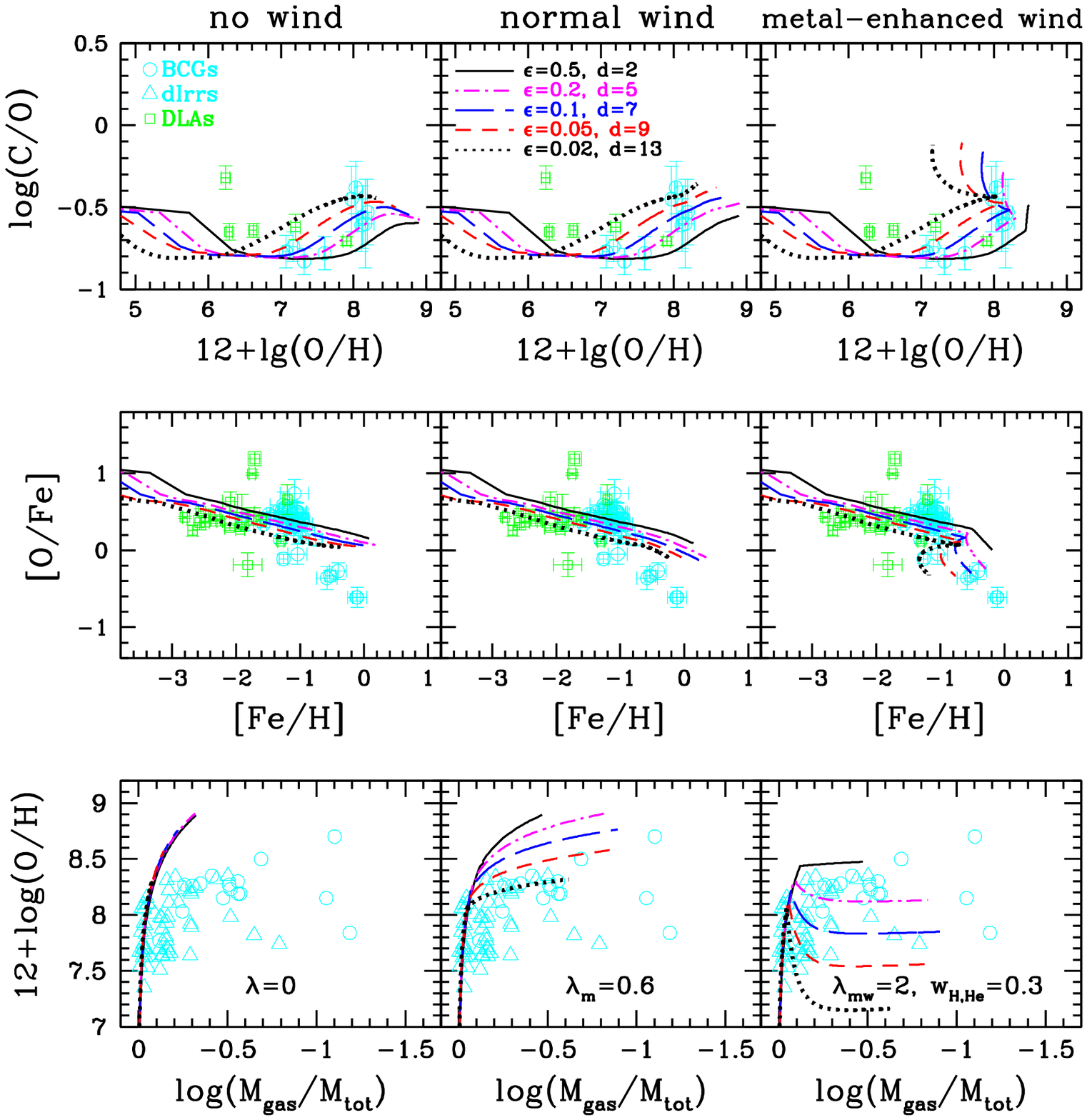}\\
  \caption{The evolutionary track as predicted by models with
continuous star formation. From top to bottom panels are log(C/O)
vs. 12+log(O/H), [O/Fe] vs. [Fe/H], and the $mu-Z$ relation; from
left to right columns are for the cases of no wind, normal wind, and
metal-enhanced wind ($w_{H,He}=0.3$). Models with different SFEs are
shown in each case: {\it black-solid lines} for $\epsilon=0.5
~(d=2)$, {\it magenta-dash-dot lines} for $\epsilon=0.2~(d=5)$, {\it
blue-long-dash lines} for $\epsilon=0.1~(d=7)$, {\it red-dash lines}
for $\epsilon=0.05~(d=9)$, and {\it black-dot lines} for
$\epsilon=0.02~(d=13)$.
 The total infall masses are assumed to be $M_{inf}=10^9$\ms in all these models.
The observational data are the same as in
Fig.~\ref{Fig:obsmuZ} and Fig.~\ref{Fig:obsabund}, BCDs, dIrrs, and
DLAs are plotted in {\it cyan open circles, cyan open triangles} and
{\it green open squares} respectively.}
  \label{Fig:contiSFZmu}
\end{figure*}

As we have shown in the last section, highly enriched or very strong
winds will reduce the galactic oxygen abundance to a very low value
during the interburst time which has not been confirmed from the
observational point of view. Thus, in our best models $w_{\rm
H,He}=0.3$ and $\lambda_{mw}=0.8$ are assumed. Different numbers of
bursts are examined for $M_{inf}=10^9$\ms~galaxies, and the best fit
models are shown in Fig.~\ref{Fig:bestabund} and
Fig.~\ref{Fig:bestmuY}. In these models, the same SFEs
($\epsilon=0.5$) and  burst durations (d=0.3) are assumed, and the
details of each model are listed in Table.~\ref{Tab:best}.

We show the abundance ratios of different elements relative to
oxygen or iron as predicted by our best models for
$M_{inf}=10^9$\ms~ in Fig.~\ref{Fig:bestabund}. The evolutionary
tracks of these models can pass through the DLA data at early time,
and cover the regions where most of BCDs have been observed.

The evolutionary tracks of the $\mu-Z$ relation (upper panel) and
$Z-Y$ relation (lower panel) predicted by our best models are shown
in Fig.~\ref{Fig:bestmuY}. After the wind develops, more are the
bursts that a galaxy suffers, more are the oscillations in the
evolutionary tracks, and these tracks pass through most of the data,
thus explaining the observed spread. On the other hand, in the panel
showing $Y$ vs. (O/H) helium keeps increasing while O is oscillating
and this is due to the fact that this element is produced also
during the interburst periods and lost less effectively than O. This
implies again that we should use the lower envelop of the
observational data to derive the primordial abundance of helium.

In Fig.~\ref{Fig:bestSNIa}, we show the evolution of the Type Ia
supernova rates predicted by our best models. The rates in this
figure are normalized to the galaxy stellar mass at that time, i.e.,
expressed in number of SNe per century and per $10^{10}$\ms. The
peaks are always associated with the star formation periods, but SNe
Ia also explode during the interburst times. Compared to the
observed value quoted for the Irr galaxies, $0.77^{+0.42}_{-0.31}$
per century and per $10^{10}$\ms \citep{Mannucci05}, our model
predicts a very high normalized SN Ia rate during the first star
formation burst, but $\sim1-2$ Gyr later, the rate decays, and
becomes comparable to the Irr galaxies in the following SF periods.
In these best models, the maximum of SN Ia rate by number varies
between 0.022 and 0.024 per century and the present value varies
between 0.0012 and 0.0016 per century (see Table~\ref{Tab:SNIa}).
\citet{Sullivan06} have studied the relation between SN Ia rate and
the stellar mass of the host galaxy in the redshift range
$0.2-0.75$, and they found the SN Ia rate is less than 0.01 per
century for star-forming galaxies whose stellar mass around
$10^8$\ms, in agreement with our predictions.

In order to reproduce the mass-metallicity relation, we have studied
galaxies of different masses, with more massive galaxies preferring
longer star formation bursts, $d=0.9$ Gyr for $M_{inf}=10^{10}$\ms~
and $d=0.1$ Gyr for $M_{inf}=10^8$\ms~ (see details in
Table~\ref{Tab:best}). The mass-metallicity relations predicted by
our best models with different numbers of bursts are shown in
Fig.~\ref{Fig:bestMsZ}, and they are consistent with the observed
one considering the scatter of the data, especially in the cases
where the wind develops in the whole galactic mass range (i.e.,
$n\ge5$).

Considering that dwarf galaxies could be in different evolutionary
stages and/or have different ages (measured from the beginning of
star formation), we show that our best models -- a series of uniform
models with same $M_{inf}$, $w_{\rm H,He}$, $\lambda_{mw}$,
$\epsilon$, $d$ but different $n$ and suitable $t$ -- can well
reproduce the spread in the observations. {\it Therefore, our
preferred galaxy formation scenario for these galaxies is the
following: they should have accreted a lot of primordial gas at
their early stages, and formed stars through several short star
bursts ($d\sim0.3$ Gyr for $M_{inf}=10^9$\ms), with more massive
galaxies suffering longer star formation bursts. However, it is
likely that the gas escape from the potential well of the galaxy
when enough energy from SN explosions is accumulated, and this wind
should be metal-enhanced ($w_{\rm H,He}\sim0.3$). The wind rate
should be proportional to the gas mass at that time, and the wind
efficiency should be $\lambda_{mw}<1$.}

DLAs could be the progenitors of dwarf irregular  galaxies, as
already suggested by \cite{Matteucci97} and \cite{Calura03}. In our
best models, we adopted a fixed duration for each burst with the
purpose of changing the parameters as less as possible. The
evolutionary tracks of our best models pass through the DLA data,
but could not explain the scatter. However, if we reduce the
duration of the first SF burst to $d=0.01\sim0.1$ Gyr, by taking
model M9b3 as an example, the models can explain the scatter in the
abundance ratios of DLAs much better (see Fig.~\ref{Fig:DLAabund}
and Table~\ref{Tab:best} for details of these models). Therefore, we
confirm that the DLA systems could be progenitors of local dwarf
galaxies.

\subsection{Continuous star formation}

By using a spectrophotometic model coupled with a chemical evolution
model, \citet{Legrand00a} demonstrated that a continuous but very
mild star formation rate (SFE as low as $10^{-3}$ \ms Gyr$^{-1}$) is
able to reproduce the main properties of IZw 18, one of the most
metal-poor BCDs we know. Therefore, in this section we are showing
model results obtained with continuous and mild SF.

We run the same models as in \citet{Legrand00a}, one model with a
mild continuous SF ($\epsilon=0.001$ Gyr$^{-1}$) only, and the other
one with a mild continuous SF ($\epsilon=0.001$ Gyr$^{-1}$) and a
current burst ($SFR=0.023$\ms yr$^{-1}$, i.e., $\epsilon=0.88$
Gyr$^{-1}$ if the observed $M_{\rm HI}=2.6\times10^7$\ms, during the
last 20 Myrs). The abundance ratios of different elements as
predicted by these two models are shown in
Fig.~\ref{Fig:Legrandabund}. As expected, such a low continuous SF
predicts a too low oxygen abundance which could not explain the
majority of local BCDs. In addition, by examining the evolutionary
tracks in  the low metallicity range, one sees that DLAs cannot be
the progenitors of such galaxies.

Therefore, we developed other models with higher SFEs and different
SF duration. Since both dIrrs and BCDs harbor recent SF activities,
we assumed that the SF is still going on at the present day, and
therefore a short duration $d$ implies a late starting time
($t_G-d$) of SF.

We have examined the models with different SFHs by varying the SFE,
duration, and infall timescale. The results show that only models
with different SFEs are able to explain the scatters in the
abundance ratios. In Fig.~\ref{Fig:contiSFZmu}, models with
different SFEs for the cases of no wind (left column), normal wind
(middle column), and metal-enhanced wind ($w_{\rm H,He}=0.3$, right
column) are shown. In these models, to avoid reaching a too high
metallicity at the present time, a  duration shorter than the age of
the universe is assumed for the highest SFE case. A short infall
timescale is adopted ($\tau=1$ Gyr) here, because it does not effect
the evolutionary track too much in the continuous SF scenario,
especially when the SF starts late.

There are several conclusions that can be drawn from
Fig.~\ref{Fig:contiSFZmu}:
\begin{enumerate}
\item In the case of no wind, although models with different
$\epsilon$ could partly explain the scatters in the abundance
ratios, they cannot reproduce the $\mu-Z$ relation. As we can see
from the bottom left panel of Fig.~\ref{Fig:contiSFZmu}, all the
evolutionary tracks are overlapping, the same as in the bursting SF
scenario without galactic wind;
\item When the wind is included in the model, the gas fraction
decreases with time  (the bottom middle and right panels of
Fig.~\ref{Fig:contiSFZmu}). We adopt relatively high wind
efficiencies in the continuous SF scenario.
When we compare the models with the same SFE in both normal wind and
metal-enhanced wind cases, we can see that their gas fractions reach
almost the same values at the present time since their gas loss
rates are comparable ($\lambda_w=w_{\rm H,He}\lambda_{mw}=0.6$);
however, their metals show different behaviors. The oxygen abundance
decreases dramatically in the metal-enhanced wind case, thus
explaining the scatter in $\mu-Z$ relation much better. In the
meantime, the metal enhanced wind model can fit the data relative to
abundance ratios much better (upper and middle panels of right
column in Fig.~\ref{Fig:contiSFZmu});
\item As we have demonstrated before, galaxies with long continuous
but mild SF (e.g., $\epsilon=0.01,~d=13$), which could be the case
for some  BCDs,  as suggested by the previous work (e.g.,
\citealt{Legrand00a, Legrand00b}), should not be the majority due to
the too low predicted oxygen abundances. Their evolutionary tracks
relative to abundance ratios also do not fit the data. In addition,
DLAs cannot be the progenitors of these objects when abundance
ratios are examined, because of the very high predicted values of
N/O, N/Fe and N/Si at low metallicity (see
Fig.~\ref{Fig:Legrandabund}), at variance with the properties of
DLAs.
\item In the continuous SF scenario, the optimal models should
have $\epsilon=0.05-0.2$, a duration of SF shorter than 13 Gyr, and
metal-enhanced winds should occur, a situation similar to a long
starburst scenario. Moreover, DLAs can be the progenitors of these
galaxies.
\end{enumerate}

\section{Discussion and Conclusions }

We have discussed in detail the chemical evolution of late-type
dwarf galaxies (dwarf irregular and blue compact galaxies) and used
the most recent data as a comparison. We have taken into account the
measured abundances of single elements (He,C,  N, O, S, Si and Fe)
as well as the gas masses. We have assumed that the late-type dwarf
galaxies form by cold gas accretion and we run models for different
accreted  baryonic masses ($10^{8}, 10^{9}$ and $10^{10}
M_{\odot}$). We have tested both bursting and continuous star
formation. We then have studied in detail the development of
galactic winds by assuming a dark matter halo which is assumed 10
times the amount of of the baryonic mass and feedback from SNe and
stellar winds. Our main conclusions are:

\begin{enumerate}
\item Galactic winds are necessary to reproduce the main properties
of late-type dwarf galaxies and they should be metal-enhanced,
namely metals should be carried away preferentially, in agreement
with previous dynamical work (e.g. \citealt{Mac99, Recchi01}). The
rate of gas loss was assumed to be proportional to the amount of gas
present at the time of the wind, which is equivalent to say that is
proportional to the SFR, in agreement with previous papers and
observational evidence \citep{Martin05, Rupke05, Chen10}. The wind
efficiency $\lambda_{mw}$ should be relatively low ($\lambda_{mw}
\sim 0.8$ in bursting SF scenario, and $\lambda_{mw} \sim 2$ in
continuous SF scenario) compared to what is assumed for dwarf
spheroidals ($\sim$ 6 to 15) where the wind should carry away all
the  residual gas (e.g. \citealt{Lanfranchi03}).

\item  Both bursting and
continuous SF scenarios have been examined. In the case of bursting
SF, the number of bursts could be no more than $\sim10$ and the star
formation efficiency could be $\sim 0.5$ Gyr$^{-1}$. Galaxies with a
long continuous but mild SF ($\epsilon\lesssim0.02$ Gyr$^{-1}$,
$d\simeq13$ Gyr) should not be the majority, whereas galaxies with
higher SFE (0.05 Gyr$^{-1}\lesssim\epsilon\lesssim$0.2 Gyr$^{-1}$)
and shorter SF duration (5 Gyr $\lesssim d\lesssim$ 9 Gyr) are still
acceptable.

\item Models with a different number of bursts and/or different star
formation efficiency and/or different burst duration, coupled with
metal enhanced winds, can reproduce at best the spread observed in
the abundances versus fractionary mass of gas and abundance ratios
versus abundances. Normal winds, where all the gas and metals are
lost at the same rate, should be rejected since they subtract too
much gas.

\item We studied the $M-Z$ relation for late-type dwarf galaxies
and we showed that in order to reproduce such a relation one should
assume that different galaxies suffered an increasing amount of star
formation (efficiency, number of bursts, duration of bursts) with
galactic mass, even without galactic winds. However, metal-enhanced
wind are necessary to reproduce all the other features. In the case
of the $M-Z$ relation,
a metal-enhanced wind efficiency increasing with galactic mass can
very well reproduce the data, leaving all the other parameters to be
the same irrespective of the galactic mass. On the other hand,
normal winds occurring at a different efficiency in different
galaxies cannot reproduce the $M-Z$ relation unless the primordial
gas is infalling continuously (i.e., a long infall timescale), or
the SFE or the number of bursts increase with galactic mass.

\item A comparison of our best model predictions with data for
DLAs has shown that these objects can well be the progenitors of
local dIrrs and BCDs, in agreement with previous papers.

\item In this work, our models can reproduce the chemical
properties of both dIrrs and BCDs. To distinguish between these two
types of galaxies, the photometric and spectral information should
also be taken into account.

\end{enumerate}

\acknowledgements  J.Y. thanks the hospitality of the Department of
Physics of the University of Trieste where this work was
accomplished. J.Y. and F.M. acknowledge the financial support from
PRIN2007 from Italian Ministry of Research, Prot. no.
2007JJC53X-001. J.Y. also thanks the financial support from the
National Science Foundation of China No.10573028, the Key Project
No.10833005, the Group Innovation Project No.10821302, 973 program
No. 2007CB815402, and the Knowledge Innovation Program of the
Chinese Academy of Sciences No. Y090761009. Finally, we thank the
referee, Leticia Carigi, for carefully reading the manuscript and
giving us very useful suggestions.

{}


\begin{thebibliography}{}

\bibitem[Abate et al. (2008)]{Abate08}Abate, A., Bridle, S., Teodoro, L. F. A., Warren, M. S., \& Hendry, M. 2008, MNRAS, 389, 1739 
\bibitem[Aloisi et al. (2007)]{Aloisi07}Aloisi, A., Clementini, G., Tosi, M., Annibali, F., Contreras, R., Fiorentino, G., Mack, J., Marconi, M., et al. 2007, ApJ, 667, L151
\bibitem[Bertin et al. (1992)]{Bertin92}Bertin, G., Saglia, R. P., \& Stiavelli, M. 1992, ApJ, 384, 423
\bibitem[Bomans et al. (1997)]{Bomans97}Bomans, D. J. Chu, Y.-H., \& Hopp, U. 1997, ApJ, 113, 1678
\bibitem[Bradamante et al. (1998)]{Bradamante98}Bradamante F., Matteucci F., \& D'Ercole A. 1998, A\&A, 337, 338
\bibitem[Brodie \& Huchra (1991)]{Brodie91}Brodie, J. P., \& Huchra, J. P. 1991, ApJ, 379, 157
\bibitem[Calura et al. (2003)]{Calura03}Calura, F., Matteucci, F., \& Vladilo, G. 2003, MNRAS, 340, 59
\bibitem[Carigi et al. (1999)]{Carigi99}Carigi, L., Col{\'i}n, P., \& Peimbert, M. 1999, ApJ, 514, 787
\bibitem[Centuri\'on et al. (2003)]{Centurion03}Centuri\'on, M., Molaro, P., Vladilo, G., P\'eroux, C., Levshakov, S. A., \& D'Odorico, V. 2003, A\&A, 403, 55
\bibitem[Chen et al. (2010)]{Chen10}Chen, Y. M., Tremonti, C. A., Heckman, T. M., Kauffmann, G., Weiner, B. J., Brinchmann, J., Wang, J. 2010, arXiv:1003.5425
\bibitem[Dekel \& Silk (1986)]{Dekel86}Dekel, A. \& Silk, J. 1986, ApJ, 303, 39
\bibitem[Dessauges-Zavadsky et al. (2004)]{Dessauges04}Dessauges-Zavadsky, M., Calura, F., Prochaska, J. X., D'Odorico, S., \& Matteucci, F. 2004, A\&A, 416, 79
\bibitem[Dessauges-Zavadsky et al. (2007)]{Dessauges07}Dessauges-Zavadsky, M., Calura, F., Prochaska, J. X., D'Odorico, S., \& Matteucci, F. 2007, A\&A, 470, 431
\bibitem[Dessauges-Zavadsky et al. (2001)]{Dessauges01}Dessauges-Zavadsky, M., D'Odorico, S., McMahon, R. G., Molaro, P., Ledoux, C., P{\'e}roux, C., Storrie-Lombardi, L. J. 2001, A\&A, 370, 426
\bibitem[Dessauges-Zavadsky et al. (2006)]{Dessauges06}Dessauges-Zavadsky, M., Prochaska, J. X., D'Odorico, S., Calura, F., \& Matteucci, F. 2006, A\&A, 445, 93
\bibitem[De Young \& Gallagher (1990)]{DeYoung90}De Young, D. S. \& Gallagher, J. S. 1990, ApJ, 356, 15
\bibitem[D'Odorico \& Molaro (2004)]{DOdorico04}D'Odorico, V., \& Molaro, P. 2004, A\&A, 415, 879
\bibitem[Ekta \& Chengalur (2010)]{Ekta10}Ekta, B. \& Chengalur, J.~N. 2010, MNRAS, 406, 1238
\bibitem[Ellison \& Lopez (2001)]{Ellison01a}Ellison, S. L. \& Lopez, S. 2001, A\&A, 380, 117
\bibitem[Ellison et al. (2001)]{Ellison01b}Ellison, S. L., Pettini, M., Steidel, C. C., \& Shapley, A. E. 2001, ApJ, 549, 770
\bibitem[Fujita et al. (2003)]{Fujita03}Fujita, A., Martin, C. L., Mac Low, M.-M., \& Abel, T. 2003, ApJ, 599, 50
\bibitem[Ferrara \& Tolstoy (2000)]{Ferrara00}Ferrara, A. \& Tolstoy E. 2000, MNRAS, 313, 291, 309
\bibitem[Garnett (2002)]{Garnett02} Garnett, D. R. 2002, ApJ, 581, 1019
\bibitem[Garnett \& Shields (1987)]{Garnett87}Garnett, D. R. \& Shields, G. A. 1987, ApJ, 317, 82
\bibitem[Grebel (2001)]{Grebel01}Grebel, E. K. 2001, ASPC, 239, 280
\bibitem[Guseva et al. (2001)]{Guseva01}Guseva, N. G., Izotov, Y. I., Papaderos, P., Chaffee, F. H., Foltz, C. B., Green, R. F., Thuan, T. X., Fricke, K. J., \& Noeske, K. G. 2001, A\&A, 378, 756
\bibitem[Guseva et al. (2003a)]{Guseva03a}Guseva, N. G., Papaderos, P., Izotov, Y. I., Green, R. F., Fricke, K. J., Thuan, T. X., \& Noeske, K. G. 2003a, A\&A, 407, 91
\bibitem[Guseva et al. (2003b)]{Guseva03b}Guseva, N. G., Papaderos, P., Izotov, Y. I., Green, R. F., Fricke, K. J., Thuan, T. X., \& Noeske, K. G. 2003b, A\&A, 407, 105
\bibitem[Henry et al. (2000)]{Henry00}Henry, R. B. C., Edmunds, M. G., \& K\"oppen, J. 2000, ApJ, 541, 660
\bibitem[Henry \& Prochaska(2007)]{Henry07}Henry, R. B. C. \& Prochaska, J. X. 2007, PASP, 119, 962
\bibitem[Izotov et al. (1999)]{Izotov99b}Izotov, Y. I., Chaffee, F. H., Foltz, C. B., Green, R. F., Guseva, N. G., \& Thuan, T. X. 1999, ApJ, 527, 757
\bibitem[Izotov et al. (2001a)]{Izotov01a}Izotov, Y. I., Chaffee, F. H., \& Green, R. F. 2001a, ApJ, 562, 727
\bibitem[Izotov et al. (2001b)]{Izotov01b}Izotov, Y. I., Chaffee, F. H., \& Schaerer, D. 2001b, A\&A, 378, L45
\bibitem[Izotov et al. (2006)]{Izotov06}Izotov, Y. I., Schaerer, D., Blecha, A., Royer, F., Guseva, N. G., \& North, P. 2006, A\&A, 459, 71
\bibitem[Izotov \& Thuan (1998a)]{Izotov98a}Izotov, Y. I. \& Thuan, T. X. 1998a, ApJ, 497, 227
\bibitem[Izotov \& Thuan (1998b)]{Izotov98b}Izotov, Y. I. \& Thuan, T. X. 1998b, ApJ, 500, 188
\bibitem[Izotov \& Thuan (1999)]{Izotov99}Izotov, Y. I. \& Thuan, T. X. 1999, ApJ, 511, 639
\bibitem[Izotov \& Thuan (2004a)]{Izotov04a}Izotov, Y. I. \& Thuan, T. X. 2004a, ApJ, 602, 200
\bibitem[Izotov \& Thuan (2004b)]{Izotov04b}Izotov, Y. I. \& Thuan, T. X. 2004b, ApJ, 616, 768
\bibitem[Izotov et al. (1997)]{Izotov97}Izotov, Y. I., Thuan, T. X., \& Lipovetsky, V. A. 1997, ApJS, 108, 1
\bibitem[Jenkins (2009)]{Jenkins09}Jenkins, E. B. 2009, ApJ, 700, 1299
\bibitem[Karachentsev et al. (2004)]{Karachentsev04}Karachentsev, I. D., Karachentseva, V. E., Huchtmeier, W. K., \& Makarov, D. I. 2004, AJ, 127, 2031
\bibitem[Kauffmann et al. (1993)]{Kauffmann93}Kauffmann, G., White, S. D. M., \& Guiderdoni, B. 1993, MNRAS, 264, 201
\bibitem[Kulkarni et al. (1996)]{Kulkarni96}Kulkarni, V. P., Huang, K., Green, R. F., Bechtold, J., Welty, D. E. \& York, D. G. 1996, MNRAS, 279, 197
\bibitem[Kunth et al. (1988)]{Kunth88}Kunth, D., Maurogordato, S., \& Vigroux, L. 1988, A\&A, 204, 10
\bibitem[Lamareille et al. (2004)]{Lamareille04}Lamareille, F., Mouhcine, M., Contini, T., Lewis, I., \& Maddox, S. 2004, MNRAS, 350, 396
\bibitem[Lanfranchi \& Matteucci (2003)]{Lanfranchi03} Lanfranchi G. A., \& Matteucci F. 2003, MNRAS, 345, 71
\bibitem[Larson (1974)]{Larson74}Larson, R. 1974, MNRAS, 169, 229
\bibitem[Lee et al. (2003a)]{Lee03a}Lee, H., McCall, M. L., Kingsburgh, R., Ross, R., \& Stevenson, C. C. 2003a, AJ, 125, 146
\bibitem[Lee et al. (2003b)]{Lee03b}Lee, H., McCall, M. L., \& Richer, M. G. 2003b, AJ, 125, 2975
\bibitem[Lee et al. (2006)]{Lee06}Lee, H., Skillman, E. D., Cannon, J. M., Jackson, D. C., Gehrz, R. D., Polomski, E. F., \& Woodward, C. E. 2006, ApJ, 647, 970
\bibitem[J.C. Lee et al. (2004)]{Lee04}Lee, J. C., Salzer, J. J., \& Melbourne, J. 2004, ApJ, 616, 752
\bibitem[Ledoux et al. (2006)]{Ledoux06}Ledoux, C., Petitjean, P., Fynbo, J. P. U., M\o ller, P., \& Srianand, R. 2006, A\&A, 457, 71
\bibitem[Ledoux et al. (2003)]{Ledoux03}Ledoux, C., Petitjean, P., \& Srianand, R. 2003, MNRAS, 346, 209
\bibitem[Legrand (2000)]{Legrand00a}Legrand, F. 2000, A\&A, 354, 504
\bibitem[Legrand et al. (2000)]{Legrand00b}Legrand, F., Kunth, D., Roy, J.-R., Mas-Hesse, J. M., \& Walsh, J. R. 2000, A\&A, 355, 891
\bibitem[Lequeux et al. (1979)]{Lequeux79}Lequeux, J., Peimbert, M., Rayo, J. F., Serrano, A., \& Torres-Peimbert, S. 1979, A\&A, 80, 155
\bibitem[Levshakov et al. (2002)]{Levshakov02}Levshakov, S. A., Dessauges-Zavadsky, M., D'Odorico, S., Molaro, P. 2002, ApJ, 565, 696
\bibitem[Lipovetsky et al. (1999)]{Lipovetsky99}Lipovetsky, V. A., Chaffee, F. H., Izotov, Y. I., Foltz, C. B., Kniazev, A. Y., \& Hopp, U. 1999, ApJ, 519, 177
\bibitem[Lopez \& Ellison (2003)]{Lopez03}Lopez, S. \& Ellison, S. L. 2003, A\&A, 403, 573
\bibitem[Lopez et al. (2002)]{Lopez02}Lopez, S., Reimers, D., D'Odorico, S., \& Prochaska, J. X. 2002, A\&A, 385, 778.
\bibitem[Lu et al. (1996)]{Lu96} Lu, L., Sargent, W. L. W., Barlow, T. A., Churchill, C. W., \& Vogt, S. S. 1996, ApJS, 107, 475
\bibitem[Luridiana et al. (2003)]{Luridiana03}Luridiana, V., Peimbert, A., Peimbert, M., \& Cervi\~no, M. 2003, ApJ, 592, 846
\bibitem[Mac Low \& Ferrara (1999)]{Mac99} Mac Low, M.-M. \& Ferrara, A. 1999, ApJ, 513, 142
\bibitem[Mannucci et al. (2005)]{Mannucci05}Mannucci F., Della valle M., Panagia N., Cappellaro E., Cresci G., Maiolino R., Petrosian A., \& M. Turatto. 2005, A\&A, 433, 807
\bibitem[Marconi et al. (1994)]{Marconi94}Marconi, G., Matteucci, F., \& Tosi, M. 1994, MNRAS, 270, 35
\bibitem[Martin (1996)]{Martin96}Martin, C. L. 1996, ApJ, 465, 680
\bibitem[Martin (2005)]{Martin05}Martin, C. L. 2005, ApJ, 621, 227
\bibitem[Martin et al. (2002)]{Martin02}Martin, C. L., Kobulnicky, H. A., \& Heckman T. M. 2002, ApJ, 574 663
\bibitem[Mart{\'{\i}}n-Manj{\'o}n et al. (2008)]{Martin08}Mart{\'{\i}}n-Manj{\'o}n, M.~L., Moll{\'a}, M., D{\'{\i}}az, A.~I., \& Terlevich, R. 2008, MNRAS, 385, 854.
\bibitem[Mart{\'{\i}}n-Manj{\'o}n et al. (2009)]{Martin09}Mart{\'{\i}}n-Manj{\'o}n, M.~L., Moll{\'a}, M., D{\'{\i}}az, A.~I., \& Terlevich, R. 2009, arXiv0901.1186
\bibitem[Mateo (1998)]{Mateo98}Mateo, M. 1998, ARA\&A, 36, 435
\bibitem[Matteucci \& Chiosi (1983)]{Matteucci83}Matteucci, F. \& Chiosi, C. 1983, A\&A, 123, 121
\bibitem[Matteucci et al. (1997)]{Matteucci97}Matteucci, F., Molaro, P., \& Vladilo, G. 1997, A\&A, 321, 45
\bibitem[Matteucci \& Tosi (1985)]{Matteucci85}Matteucci, F. \& Tosi, M. 1985, MNRAS, 217, 391
\bibitem[Mendes de Oliveira et al. (2006)]{Mendes06}Mendes de Oliveira, C., Temporin, S., Cypriano, E. S., Plana, H., Amram, P., Sodr\'e, L., Jr., \& Balkowski, C. 2006, AJ, 132, 570
\bibitem[Meurer et al. (1992)]{Meurer92}Meurer, G. R., Freeman, K. C., Dopita, M. A., \& Cacciari C. 1992, AJ, 103, 60
\bibitem[Molaro et al. (2001)]{Molaro01}Molaro, P., Levshakov, S. A., D'Odorico, S., Bonifacio, P, \& Centuri\'on, M. 2001, ApJ, 549, 90
\bibitem[Noterdaeme et al. (2008)]{Noterdaeme08}Noterdaeme, P., Petitjean, P., Ledoux, C., Srianand, R., \& Ivanchik, A. 2008, A\&A, 491, 397
\bibitem[Noterdaeme et al. (2007)]{Noterdaeme07}Noterdaeme, P., Petitjean, P., Srianand, R., Ledoux, C., \& Le Petit, F. 2007, A\&A, 469, 425
\bibitem[O'Meara et al. (2006)]{OMeara06}O'Meara, J. M., Burles, S., Prochaska, J. X., Prochter, G. E., Bernstein, R. A., Burgess, K. M. 2006, ApJ, 649, L61
\bibitem[{\"O}stlin (2000)]{Ostlin00}{\"O}stlin, G. 2000, ApJ, 535, L99
\bibitem[Outram et al. (1999)]{Outram99}Outram, P. J., Chaffee, F. H., \& Carswell, R. F. 1999, MNRAS, 310, 289
\bibitem[Papaderos et al. (2006)]{Papaderos06}Papaderos, P., Guseva, N. G., Izotov, Y. I., Noeske, K. G., Thuan, T. X., \& Fricke, K. J. 2006, A\&A, 457, 45
\bibitem[Papaderos et al. (2002)]{Papaderos02}Papaderos, P., Izotov, Y. I., Thuan, T. X., Noeske, K. G., Fricke, K. J., Guseva, N. G., \& Green, R. F. 2002, A\&A, 393, 461
\bibitem[Papaderos et al. (1996)]{Papaderos96}Papaderos, P., Loose, H.-H., Thuan, T. X., \& Fricke, K. J. 1996, A\&AS, 120, 207
\bibitem[A. Peimbert (2003)]{Peimbert03}Peimbert, A. 2003, ApJ, 584, 735
\bibitem[M. Peimbert (2007)]{Peimbert07}Peimbert, M., Luridiana V., \& Peimbert A. 2007, ApJ, 666, 636
\bibitem[P\'erez-Gonz\'alez et al. (2003)]{Perez03}P\'erez-Gonz\'alez, P. G., Gil de Paz, A., Zamorano, J., Gallego, J., Alonso-Herrero, A., \& Arag\'on-Salamanca, A. 2003, MNRAS, 338, 525
\bibitem[P\'eroux et al. (2006)]{Peroux06}P\'eroux, C., Meiring, J. D., Kulkarni, V. P., Ferlet, R., Khare, P., Lauroesch, J. T., Vladilo, G., \& York, D. G. 2006, MNRAS, 372, 369
\bibitem[Petitjean et al. (2008)]{Petitjean08}Petitjean, P., Ledoux, C., \& Srianand, R. 2008, A\&A, 480, 349
\bibitem[Petitjean et al. (2000)]{Petitjean00}Petitjean, P., Srianand, R., \& Ledoux, C. 2000, A\&A, 364, L26
\bibitem[Pettini et al. (2002)]{Pettini02}Pettini, M., Ellison, S. L., Bergeron, J., \& Petitjean, P. 2002, A\&A, 391, 21
\bibitem[Pettini et al. (2008)]{Pettini08}Pettini, M., Zych, B. J., Steidel, C. C., \& Chaffee, F. H. 2008, MNRAS, 385, 2011
\bibitem[Pilyugin (1993)]{Pilyugin93}Pilyugin, L. S. 1993, A\&A, 277, 42
\bibitem[Pilyugin (2001)]{Pilyugin01}Pilyugin, L. S. 2001, A\&A, 374, 412
\bibitem[Pilyugin et al. (2004)]{Pilyugin04}Pilyugin, L. S., Vilchez, J. M., \& Contini, T., 2004, A\&A, 425, 849
\bibitem[Prochaska et al. (2008)]{Prochaska08}Prochaska, J. X., Chen, H., Wolfe, A. M., Dessauges-Zavadsky, M., \& Bloom, J. S. 2008, ApJ, 672, 59 
\bibitem[Prochaska et al. (2003)]{Prochaska03}Prochaska, J. X., Gawiser, E., Wolfe, A. M., Cooke, J., \& Gelino, D. 2003, ApJS, 147, 227
\bibitem[Prochaska et al. (2002a)]{Prochaska02a}Prochaska, J. X., Henry, R. B. C., O'Meara, J. M., Tytler, D., Wolfe, A. M., Kirkman, D., Lubin, D., \& Suzuki, N. 2002a, PASP, 114, 933
\bibitem[Prochaska et al. (2002b)]{Prochaska02b}Prochaska, J. X., Howk, J. C., O'Meara, J. M., Tytler, D., Wolfe, A. M., Kirkman, D., Lubin, D., \& Suzuki, N. 2002b, ApJ, 571, 693
\bibitem[Prochaska et al. (2007)]{Prochaska07}Prochaska, J. X., Wolfe, A. M., Howk, J. C., Gawiser, E., Burles, S. M., \& Cooke, J. 2007, ApJS, 171, 29
\bibitem[Pustilnik et al. (2004)]{Pustilnik04}Pustilnik, S. A., Pramskij, A. G., \& Kniazev, A. Y. 2004, A\&A, 425, 51
\bibitem[Recchi et al. (2001)]{Recchi01}Recchi, S., Matteucci, F. \& D'Ercole, A., 2001, MNRAS, 322, 800
\bibitem[Recchi et al. (2002)]{Recchi02}Recchi, S., Matteucci, F. \& D'Ercole, A., 2002, A\&A, 384, 799
\bibitem[Recchi et al. (2004)]{Recchi04}Recchi, S., Matteucci, F., D'Ercole, A., \& Tosi, M. 2004, A\&A, 426, 37
\bibitem[Recchi et al. (2008)]{Recchi08}Recchi, S., Spitoni, E., Matteucci, F., \& Lanfranchi, G. A. 2008, A\&A, 489, 555
\bibitem[Romano et al. (2006)]{Romano06}Romano, D., Tosi, M., \& Matteucci, F. 2006, MNRAS, 365, 759
\bibitem[Rosenberg et al. (2006)]{Rosenberg06}Rosenberg, J. L., Ashby, M. L. N., Salzer, J. J., \& Huang, J.-S. 2006, ApJ, 636, 742
\bibitem[Rupke et al. (2005)]{Rupke05}Rupke D. S., Veilleux S., \& Sanders D. B. 2005, ApJS, 160, 115
\bibitem[Salpeter (1955)]{Salpeter55} Salpeter, E. E., 1955, ApJ, 121, 161
\bibitem[Salzer et al. (2005)]{Salzer05}Salzer, J. J., Lee, J. C., Melbourne, J., Hinz, J. L., Alonso-Herrero, A., \& Jangren, A. 2005, ApJ, 624, 661
\bibitem[Saviane et al. (2008)]{Saviane08}Saviane, I., Ivanov, V. D., Held, E. V., Alloin, D., Rich, R. M., Bresolin, F., \& Rizzi, L. 2008, A\&A, 487, 901
\bibitem[Scalo (1986)]{Scalo86}Scalo, J. M. 1986, Fund. Cosmic Phys., 11, 1
\bibitem[Schmidt (1963)]{Schmidt63}Schmidt, M. 1963, ApJ, 137, 758
\bibitem[Schulte-Ladbeck et al. (2001)]{Schulte01}Schulte-Ladbeck, R. E., Hopp, U., Greggio, L., Crone, M. M., \& Drozdovsky, I. O. 2001, ApSSS, 277, 309
\bibitem[Searle \& Sargent (1972)]{Searle72}Searle, L. \& Sargent, W. L. W. 1972, ApJ, 173, 25
\bibitem[Searle et al. (1973)]{Searle73}Searle, L., Sargent, W. L. W., \& Bagnuolo, W. G. 1973, ApJ, 179, 427
\bibitem[Skillman et al. (1997)]{Skillman97}Skillman, E. D., Bomans, D. J., \& Kobulnicky, H. A. 1997, ApJ, 474, 205
\bibitem[Skillman et al. (1989)]{Skillman89}Skillman, E. D., Kennicutt, R. C., \& Hodge, P. W. 1989, ApJ, 347, 875
\bibitem[Spergel et al. (2007)]{Spergel07}Spergel, D. N., Bean, R., Dor\'e, O., Nolta, M. R., Bennett, C. L., Dunkley, J., Hinshaw, G., Jarosik, N., et al. 2007, ApJS, 170, 377
\bibitem[Srianand \& Petitjean (2001)]{Srianand01}Srianand, R. \& Petitjean, P. 2001, A\&A, 373, 816
\bibitem[Srianand et al. (2005)]{Srianand05}Srianand, R., Petitjean, P., Ledoux, C., Ferland, G., \& Shaw, G. 2005, MNRAS, 362, 549
\bibitem[Stasi{\'n}ska \& Izotov (2003)]{Stasinska03}Stasi{\'n}ska, G. \& Izotov, Y. 2003, A\&A, 397, 71
\bibitem[Staveley-Smith et al. (1992)]{Staveley92}Staveley-Smith, L., Davies, R. D., \& Kinman, T. D. 1992, MNRAS, 258, 334
\bibitem[Storrie-Lombardi \& Wolfe (2000)]{Storrie00}Storrie-Lombardi, L. J. \& Wolfe, A. M. 2000, ApJ, 543, 552
\bibitem[Sullivan et al. (2006)]{Sullivan06}Sullivan, M., Borgne, D. Le, Pritchet, C. J., Hodsman, A., Neill, J. D., Howell, D. A., Carlberg, R. G., Astier, P., et al. 2006, ApJ, 648, 868
\bibitem[Thuan (2008)]{Thuan08}Thuan, T. X. 2008, IAUS, 255, 348
\bibitem[Thuan et al. (1999)]{Thuan99}Thuan, T. X., Izotov, Y. I., \& Foltz, C. B. 1999, ApJ, 525, 105
\bibitem[Thuan et al. (1995)]{Thuan95}Thuan, T. X., Izotov, Y. I., \& Lipovetsky, V. A. 1995, ApJ, 445, 108
\bibitem[Tosi et al. (1991)]{Tosi91}Tosi, M., Greggio, L., Marconi, G., \& Focardi, P. 1991, AJ, 102, 951
\bibitem[Tremonti et al. (2004)]{Tremonti04}Tremonti, C. A., Heckman, T. M., Kauffmann, G., Brinchmann, J., Charlot, S., White, S. D. M., Seibert, M., Peng, E. W., et al. 2004, ApJ, 613, 898
\bibitem[Vaduvescu et al. (2007)]{Vaduvescu07}Vaduvescu, O., McCall, M. L., \& Richer, M. G., 2007, AJ, 134, 604
\bibitem[Vaduvescu et al. (2005)]{Vaduvescu05}Vaduvescu, O., McCall, M. L., Richer, M. G., \& Fingerhut, R. L. 2005, AJ, 130, 1593
\bibitem[Vaduvescu et al. (2006)]{Vaduvescu06}Vaduvescu, O., Richer, M. G., \& McCall, M. L. 2006, AJ, 131, 1318
\bibitem[van den Hoek \& Groenewegen (1997)]{vanHoek97} van den Hoek, L. B., \& Groenewegen, M. A. T. 1997, A\&AS, 123, 305
\bibitem[van Zee \& Haynes (2006)]{vanZee06}van Zee, L. \& Haynes, M. 2006, ApJ, 636, 214
\bibitem[van Zee et al. (1997)]{vanZee97}van Zee, L., Haynes, M. P., \& Salzer, J. J. 1997, AJ, 114, 2497
\bibitem[van Zee et al. (1998)]{vanZee98}van Zee, L., Skillman, E. D., \& Salzer, J. J. 1998, AJ, 116, 1186
\bibitem[V{\'a}zquez et al. (2003)]{Vazquez03}V{\'a}zquez, G. A., Carigi, L., \& Gonz{\'a}lez, J. J. 2003, A\&A, 400, 31
\bibitem[Vladilo (2004)]{Vladilo04}Vladilo, G. 2004, A\&A, 421, 479
\bibitem[White \& Frenk (1991)]{White91}White, S. D. M. \& Frenk C. S. 1991, ApJ, 379, 52
\bibitem[Woosley \& Weaver (1995)]{Woosley95}Woosley, S. E. \& Weaver, T. A., 1995, ApJS, 101, 181
\bibitem[Zaritsky et al. (1994)]{Zaritsky94}Zaritsky, D., Kennicutt, R. C., Jr., \& Huchra, J. P. 1994, ApJ, 420, 87
\end{thebibliography}
\end{document}